\documentclass{aiaa-tc}

\usepackage{wrapfig}
\usepackage{threeparttable}
\usepackage{dcolumn}
\newcolumntype{d}{D{.}{.}{-1}}
\usepackage{nomencl}
\makeglossary
\usepackage{subfigure}
\usepackage{subfigmat}
\usepackage{fancyvrb}
\fvset{fontsize=\footnotesize,xleftmargin=2em}
\usepackage{lettrine}
\usepackage{hyperref}

\usepackage{booktabs}
\usepackage{bm, amssymb, amsmath, array, pdfpages}
\usepackage{amsmath,amssymb,color,hyperref,graphicx,pdfsync,overpic,color,epstopdf,rotating,dashrule,float,bm}
\usepackage{float}

\usepackage{setspace}
\usepackage{comment}

\graphicspath{{Figures/}}

\title{Uncertainty Quantification for Airfoil Icing using Polynomial Chaos Expansions}

\author{%
  Anthony M. DeGennaro\thanks{Graduate Student, Mechanical and Aerospace Engineering, Engineering Quad: Olden Street, Princeton, NJ, 08540.}
  \ ,
  Clarence W. Rowley\thanks{Professor, Mechanical and Aerospace Engineering, Engineering Quad: Olden Street, Princeton, NJ, 08540.}
  \ , and
  Luigi Martinelli\thanksibid{2}\\
  {\normalsize\itshape
   Princeton University, Princeton, NJ, 08540, USA}\\
   {\normalsize \copyright~2014 by the authors. Do not distribute without permission.}
 }

 \AIAApapernumber{200?-????}
 \AIAAconference{Conference Name, Date, and Location}
 \AIAAcopyright{\AIAAcopyrightD{200?}}

\newcommand{\CLmax}{{C_L}_\text{max}}
\newcommand{\LDmax}{L/D_\text{max}}
\newcommand{\alphamax}{\alpha_\text{max}}
\def\ip<#1,#2>{\left\langle #1,#2\right\rangle}

\begin{document}

\maketitle

\begin{abstract}
  The formation and accretion of ice on the leading edge of a wing can be
  detrimental to airplane performance. Complicating this reality is the fact
  that even a small amount of uncertainty in the shape of the accreted ice may
  result in a large amount of uncertainty in aerodynamic performance metrics
  (e.g., stall angle of attack). The main focus of this work concerns using the
  techniques of Polynomial Chaos Expansions (PCE) to quantify icing uncertainty
  much more quickly than traditional methods (e.g., Monte Carlo). First, we
  present a brief survey of the literature concerning the physics of wing icing,
  with the intention of giving a certain amount of intuition for the physical
  process. Next, we give a brief overview of the background theory of
  PCE. Finally, we compare the results of Monte Carlo simulations to PCE-based
  uncertainty quantification for several different airfoil icing scenarios. The
  results are in good agreement and confirm that PCE methods are much more
  efficient for the canonical airfoil icing uncertainty quantification problem
  than Monte Carlo methods.
\end{abstract}

\section*{Nomenclature}

\begin{tabbing}
  XXXXXXXXXXXX \= \kill
  $u$ \> State vector \\
  $x, t, Z$ \> Physical space vector, time, and stochastic parameter vector \\
  $y$ \> Observable vector \\
  $\rho(Z)$ \> Probability density function (PDF) of $Z$ \\
  $\Gamma$ \> Support of Z \\
  $\phi(Z)$ \> Univariate polynomial chaos basis function of $Z$ \\
  $\Phi(Z)$ \> Multivariate polynomial chaos basis function of $Z$ \\
  $h_i$ \> Norm of the $i^{th}$ polynomial chaos basis function \\
  $\mathbb{P}_N$ \> $N^{th}$-order projection operator \\
  $R, S_r$ \> Radius and chordwise-position (aft of leading edge) of the ridge ice shape \\
  $dR, dS_r$ \> Perturbations in the ridge ice parameters $R$ and $S_r$ from the mean ridge ice shape \\
  $H, S_h$ \> Height (normal to airfoil surface) and separation distance of the horn ice shape \\
  $h, s$ \> Scaling parameters for the height $H$ and separation distance $S_h$ of the horn ice shape \\
  $\mu, \sigma$ \> Mean and standard deviation \\
  $\CLmax$ \> Maximum lift coefficient in the steady flowfield regime\\
  $\alphamax$ \> Angle of attack at $\CLmax$\\
  $L/D_{MAX}$ \> Maximum lift to drag ratio\\
  $c$ \> Chord length \\
 \end{tabbing}

\section{Introduction}

Wing icing can present a significant safety hazard to pilots. Wing ice shapes
are often sharp and jagged in profile and rough in surface texture; both of
these characteristics tend to induce flow separation at angles of attack which
are often quite mild. This can cause stall at lower angles of attack than those
to which pilots are accustomed. It is for this reason that wing icing has been
identified as the cause of 388 crashes between 1990 and 2000 \cite{landsberg}.

Equally troubling is the fact that wing icing is a process which is associated
with a high degree of uncertainty in practice. The reasons for this are fairly
intuitive: icing itself is fundamentally a complex physics problem, in which
both aerodynamics and thermodynamics are involved and highly
coupled. Additionally, wing icing codes such as NASA's LEWICE often make ``rule
of thumb'' approximations and educated guesses at the physical parameters
governing the process\cite{lewice,myers}. Lastly, it
is obvious that, in flight, pilots can never know all of the many physical
parameters which contribute to the icing process with perfect certainty, since there
is always some uncertainty in weather-related information such as the
liquid water content of the free-stream air, the ambient temperature, the
temperature of the wing surface, etc.

The main focus of this work is to describe a fast, accurate framework for
quantifying the uncertainty which arises in aerodynamic performance metrics
(e.g., stall angle of attack, max lift coefficient, etc.) as a direct result of
uncertainty in a set of parameters which govern the airfoil ice shape.

\section{Wing Icing: Background}

The consensus in the literature is that wing icing may be conveniently divided
into a few simple classifications. Each of these is briefly discussed in
turn. The interested reader is referred to the literature
for a more in-depth discussion\cite{bragg}.

\subsection{Streamwise Ice}

As the name implies, this scenario is characterized by relatively low amounts of
smooth ice accumulation which closely follow the airfoil contour
\cite{bragg}. Aerodynamically, therefore, this ice is much less detrimental to
performance than the other classifications. Hence, we do not pursue it in this
work any further.

\subsection{Ice Roughness}

This situation is characterized by the development of a region of smooth ice
near the stagnation point which transitions aft to rough ice and then to several
small, sharp, jagged ice ``feathers.'' \cite{bragg} Depending on the size of the
feathers relative to the scale of the boundary layer, the feathers can act as
flow obstacles and promote separation. Aerodynamically, the distinction between
this scenario and the more familiar ice horn is not always clear (with the
exception that ice horns are primarily a two-dimensional (2D) phenomenon, while the ice feathers
may exhibit three-dimensional (3D) effects).While ice roughness may lead to early trailing-edge
separation, it typically does not cause the large scale separation bubbles which
the more malignant horn and ridge ice do. For these reasons, we will not discuss
this situation any further in this work.

\subsection{Ridge Ice}

In this scenario, a ridge forms aft of the stagnation point, as illustrated in
Figure~\ref{fig:HornRidgeParam}(a).
This situation is particularly dangerous for two reasons. First, the ridge
formation is jagged and discontinuous, which promotes large scale separation at
relatively low angles of attack. Second, the ridge accumulates just aft of the
deicing equipment, and hence it is not possible to combat this type of ice with
traditional deicing mechanisms. The ridge profile is typically modeled as either
a forward or backward facing quarter circle round. It is predominantly a 2D
phenomenon, and the major parameters which describe the profile shape are
the radius of the quarter circle round and its position aft of the leading edge
\cite{bragg,miller}.

\subsection{Horn Ice}

Horn ice forms in icing conditions which are relatively warmer with higher
amounts of liquid water content in the free-stream. It can be understood in
terms of a division between areas near the stagnation point of the airfoil, for
which rates of ice accretion are lower, and areas aft of that, which experience
higher rates. Initially (i.e., before the horn has begun to form), this division
is due to a coupling between aerodynamics and thermodynamics: the boundary layer
near the stagnation point is laminar, while aft of that point it transitions to
a turbulent profile, and the local convective heat transfer rate is lower for
laminar boundary layers than it is for turbulent ones \cite{yamaguchi}. As time
progresses, however, surface roughness becomes the dominant driving force in the
creation of the horn. Near the stagnation point, a thin, uniform film of water
develops on the ice surface, which keeps the ice surface smooth. Aft of the
stagnation point, the ice surface is mostly dry, with water beadlets
intermittently dispersed. When these beadlets freeze, they roughen the ice
surface, and convective heat transfer is higher for rougher ice surfaces than
for smoother ones. Because convection dominates the heat transfer, it is this
division in local convective heat transfer that is thermodynamically responsible
for horn ice \cite{hansman}. Horn ice is primarily 2D. The major parameters used
to describe the profile shape in this work are the horn height (normal to the
airfoil surface) and separation distance between the horns.

\section{Polynomial Chaos}

The ultimate goal of this work is the quantification of uncertainty in
aerodynamic performance metrics, resulting from uncertainty in the
parameters that govern the ice shape (whether ridge or horn). Traditionally,
uncertainty quantification (UQ)
problems such as this have been approached through Monte Carlo methods. The main
drawback of these methods is that they are sampling-based, and as such often
require undesirably or unfeasibly large sample sizes. A relatively new
alternative to this approach which is gaining increasing popularity is to use
polynomial chaos expansions (PCE).

Weiner coined the term ``polynomial chaos'' when studying the decomposition of
Gaussian processes with Hermite polynomials. It was Ghanem and Spanos, however,
who first introduced the stochastic spectral method for systems with Gaussian
input processes using the Hermite polynomials as a basis \cite{ghanem_book}. Xiu
and Karniadakis later generalized this spectral approach to account for
non-Gaussian input processes by using orthogonal polynomials different from the
Hermite basis \cite{xiu_book}.

There are two main approaches to UQ using PCE: the stochastic Galerkin method,
and the stochastic collocation method (see Xiu\cite{xiu_book} for more information
about both). The first method can be thought of as an extension of traditional
Galerkin methods. The governing stochastic equations are orthogonally projected
onto the span of a PCE basis, and all expansions are truncated at finite
order. This results in a new system of coupled, deterministic equations which
must be solved for the modes of the solution expansion. The stochastic
collocation method, on the other hand, is essentially a discrete Galerkin
approach, whereby the integrals in the projection equations are approximated by
a quadrature rule of sufficient accuracy. This process results in the evaluation
of the governing equation at a finite number of nodes, or ``collocation
points.''

There are advantages and disadvantages to either of these approaches. The
stochastic Galerkin approach can be very difficult to implement when the
governing equations are large or complicated, since the new, coupled equations
must be first derived and then solved. The stochastic collocation method, on the
other hand, does not require the derivation of new equations, and so any
legacy codes for solving the original equations may still
be used. It should be noted, however, that the stochastic collocation method can
suffer from aliasing error if the quadrature mesh is too coarse; the stochastic
Galerkin method does not suffer from this.

In this work, we use the stochastic collocation
method for UQ. We give a brief overview of this approach below;
further details can be found in introductory references on
PCE methods\cite{ghanem_book,xiu_book}.

\subsection{Stochastic Collocation Method}

Let~$Z=(Z_1,Z_2)$ be a vector of random variables that parameterize the
uncertain quantities in the ice shape.  For instance, for ridge ice, these
parameters will be the location and height of the ridge, and for horn ice, these
will specify the horn height and separation distance.  We are interested in the
corresponding uncertainty of an aerodynamic quantity, represented by $y(Z)$.
For instance, $y(Z)$ could be the maximum lift coefficient, angle of attack at
which the maximum lift occurs, or maximum $L/D$.

In our setting, the true values of $y(Z)$ are determined by a 2D steady-state
RANS calculation.  A particular value of $Z$ specifices the boundary conditions,
and we solve the RANS equations for these boundary conditions, to determine the
resulting aerodynamic quantities.

The goal of the method is to represent $y(Z)$ in terms of some basis
functions~$\Phi_i$, writing
\begin{equation}
  \label{eq:1}
  y(Z) = \sum_{|i|=0}^N y_i \Phi_i(Z).
\end{equation}
Here, $i=(i_1,i_2)$ is a multi-index, and $|i|=i_1+i_2$.  We define an inner
product on the space of functions of the random variables by
\begin{equation}
  \label{eq:2}
  \ip<f,g> = \int_\Gamma f(Z) g(Z) \rho(Z)\,dZ,
\end{equation}
where $\rho(Z)$ denotes the probability density function of $Z$, and has support
$\Gamma$.  We assume our basis functions are orthonormal with respect to this
inner product, so that
\begin{equation}
  \label{eq:3}
  \ip<\Phi_i,\Phi_j> = \delta_{ij},
\end{equation}
where $\delta_{ij}=1$ if $i=j$, and $0$ if $i\ne j$.  The coefficients $y_i$ in
the expansion~\eqref{eq:1} may then be deterimined by taking an inner product
with $\Phi_j$: because the $\Phi_j$ are orthonormal, we have
\begin{equation}
  \label{eq:4}
  y_j = \ip<y,\Phi_j>.
\end{equation}
Note that one could also take $y(Z)$ to be a vector of several different
aerodynamic quantities of interest: in this case, the coefficients~$y_i$ in the
expansion~\eqref{eq:1} are vectors, and each component of $y_i$ is determined by
an equation such as~\eqref{eq:4}, for the corresponding component of~$y$.

In the stochastic collocation method, we approximate the inner product
in~\eqref{eq:4} by employing a quadrature rule.  In particular, the basis
functions~$\Phi_i$ are chosen to be
\begin{equation}
  \label{eq:6}
  \Phi_i(Z) = \phi_{i_1}(Z_1) \phi_{i_2}(Z_2),
\end{equation}
where $\phi_n$ is a (univariate) polynomial of degree~$n$.  In practice, the
$\{\phi_n\}$ will be a basis of orthogonal polynomials, chosen so that the
orthogonality condition~\eqref{eq:3} is satisfied.  (For instance, if $\rho(Z)$
is a Gaussian distribution, then suitably normalized Hermite polynomials satisfy
the orthogonality relation.)
To evaluate the inner product, we then choose quadrature nodes
$Z^{(k)}$, for $k=1,\ldots,Q$, with corresponding quadrature weights $w_k$, and
approximate the inner product as
\begin{equation}
  \label{eq:5}
  y_j = \ip<y,\Phi_j> \approx \sum_{k=1}^Q y(Z^{(k)}) \Phi_j(Z^{(k)}) w_k.
\end{equation}
In practice, we use a Gauss quadrature rule to deterimine $Z^{(k)}$ and
$w_k$, taking $N+1$ points for each component of $Z$ (where $N$ is the degree of
the polynomial expansion~\eqref{eq:1}), for a total of $Q=(N+1)^2$
nodes and weights. This procedure gives exact results if the integrand is any
polynomial of degree $2N+1$ or less in any of the components of~$Z$.  The
resulting model interpolates the true solution between the nodes.

\subsection{Statistical Information}

If we have a sufficiently accurate PC expansion for the observable, defined as
in \eqref{eq:1}, then we may retrieve statistical moments through a few
simple post-processing steps. For example, the mean is approximated by the
expected value of the PC expansion: noting that $\Phi_0=1$, we have
\begin{equation}
  \mu = \mathbb{E}[y] = \int_{\Gamma} y\rho(Z)\,dZ = \ip<y, \Phi_0> = y_0.
\end{equation}
Similarly, the variance can be approximated as
\begin{equation}
\sigma^2 = \mathbb{E}[(y-\mu)^2] = \ip<y-y_0,y-y_0> = \sum_{|i|=1}^N y_i^2.
\end{equation}

\subsection{Overview of PCE Algorithm for Quantifying Uncertainty}

Putting together the results of this section, we can outline a simple set
of steps which describes how to implement a PCE-based uncertainty
quantification, based on the stochastic collocation method:

\begin{enumerate}
\item {\it Choose polynomial chaos basis.} The first step is to choose the
  polynomial chaos basis which is optimal for the chosen application. Here,
  ``optimal'' is usually defined in terms of spectral convergence: we wish to
  select that basis which best represents the input and output probability
  density functions (PDFs) using the fewest number of modes.
\item{\it Determine collocation points and weights.} Here, we wish to determine
  the mesh of discrete collocation points and weights which are required to
  evaluate \eqref{eq:5}. The procedure for selecting the collocation
  nodes/weights is a straightforward application of Gaussian quadrature
  techniques.
\item{\it Solve governing equations at collocation points.} We obtain
  the values of $y(Z)$ at each of the collocation points by solving
  the 2D steady-state RANS equations at each of the collocation points.
\item{\it Obtain the approximate PCE model.} This is done by solving
  \eqref{eq:5}. Any desired statistical information follows through
  simple post-processing.
\end{enumerate}

\section{Application to Airfoil Icing}

The PCE stochastic collocation method presented in the previous section is
applied to the ridge and horn ice problems in this section, with the goal of
quantifying uncertainty in airfoil aerodynamic performance metrics, such as max
lift coefficient and stall angle of attack. The results are compared to Monte
Carlo simulations.

We consider two classes of airfoil icing: ridge and horn. Both of these icing
cases are modeled as 2D phenomena. The airfoil profile used is the NACA 63A213
at a Reynolds number of $Re = 4.5 \times 10^5$ and Mach number of $M = 0.21$;
these conditions are chosen in agreement with a those used in a paper which
investigated simulated ice accretions using LEWICE \cite{papadakis}.

\begin{figure}[htb]
\centering
\subfigure[Ridge parameterization.]{\includegraphics[width=0.4\textwidth]{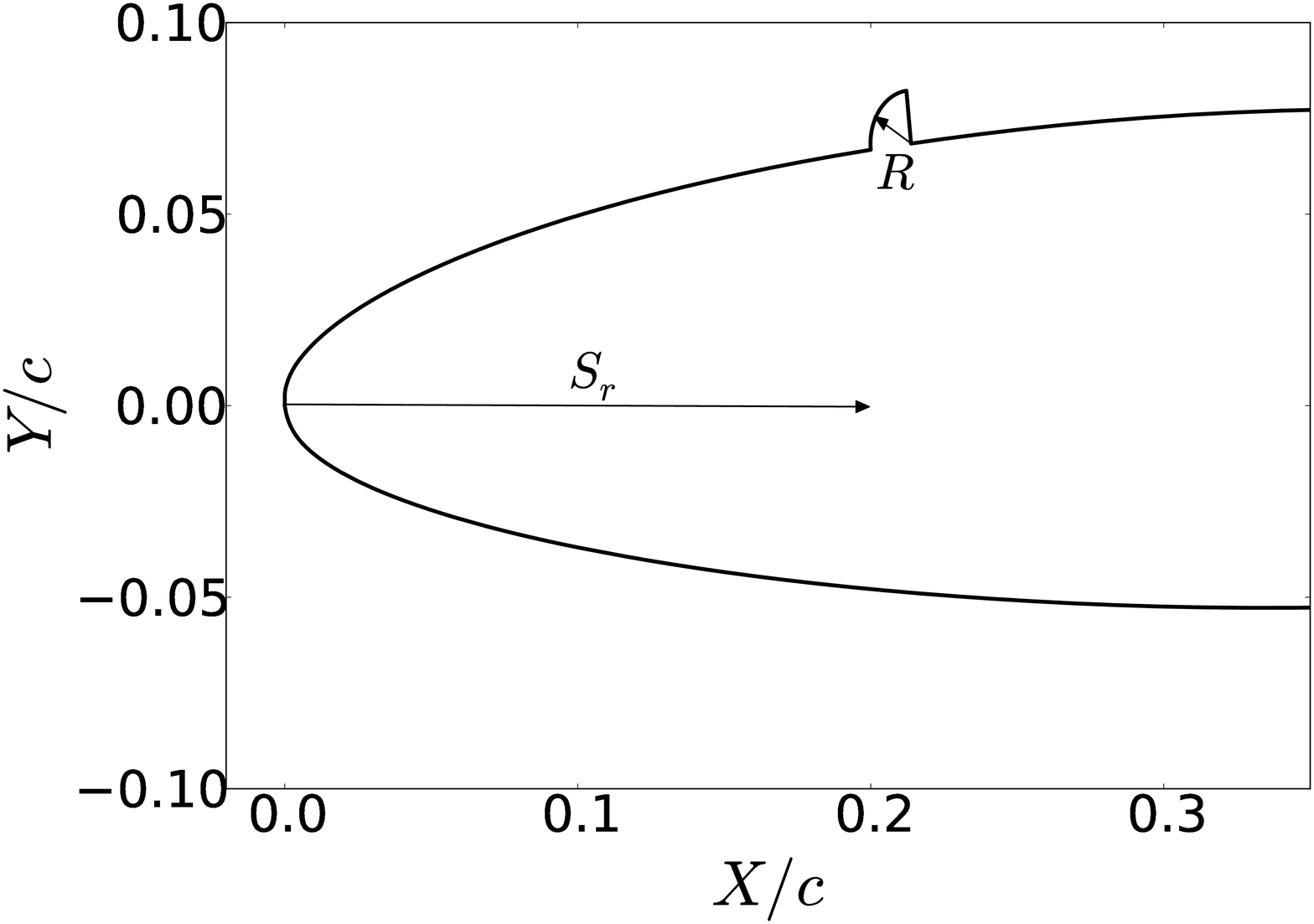}}
\subfigure[Horn parameterization.]{\includegraphics[width=0.4\textwidth]{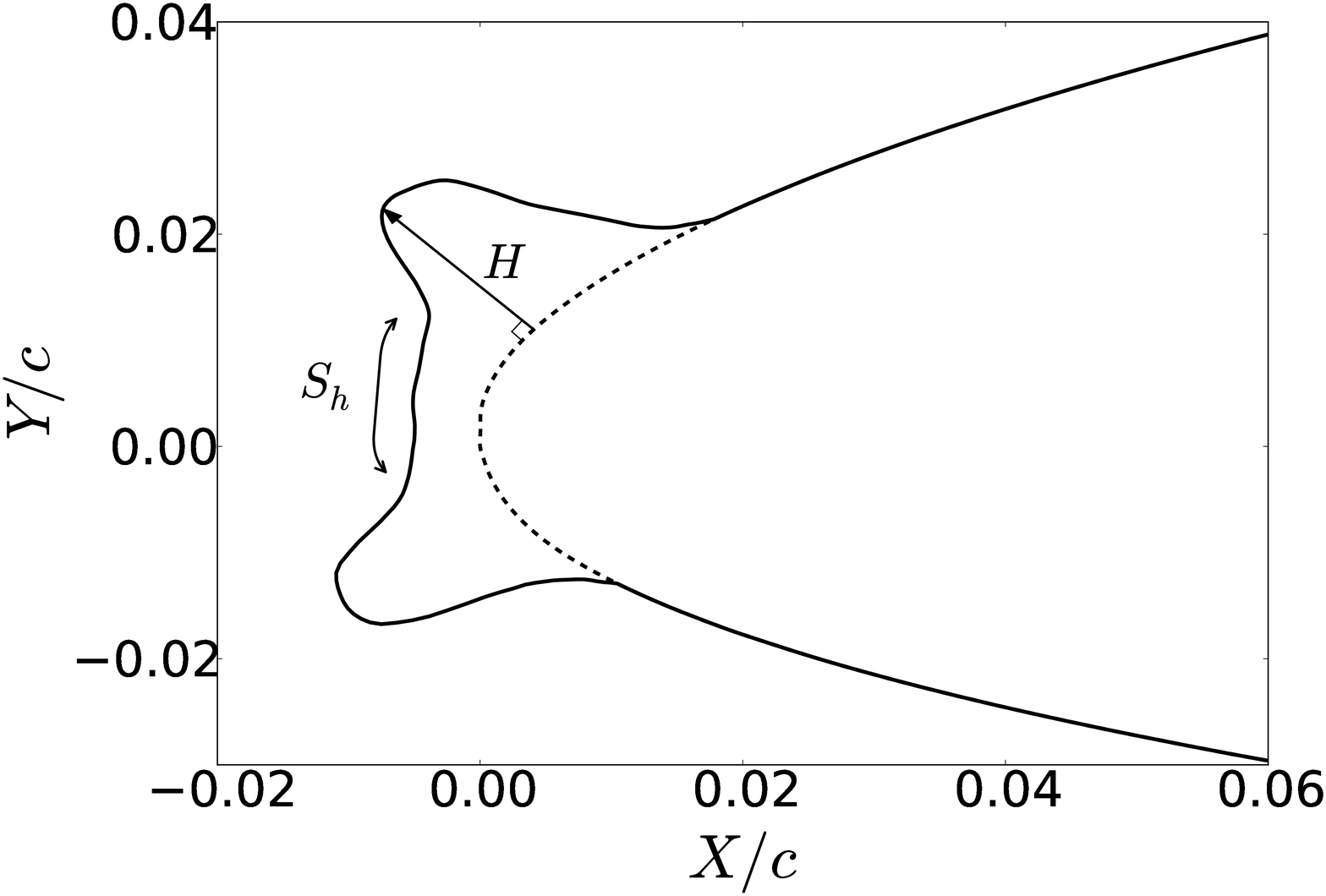}}
\caption{Parameterization convention for the ridge and horn ice scenarios. The figure at the right represents the mean horn ice shape used in this work, obtained from Papadakis\cite{papadakis}.}
\label{fig:HornRidgeParam}
\end{figure}

We model ridge ice shapes as backward-facing quarter circle rounds, which are
parameterized by the radius $R$ of the round and the location $S_r$ aft of the
airfoil leading edge (see Fig.~\ref{fig:HornRidgeParam}). The radius $R$ governs
uncertainty in the size of the ridge ice, while the position $S_r$ governs
uncertainty in where the ridge forms aft of the deicing boot.

In all cases, we specify the uncertainty in $R$ and $S_r$ as a bivariate
Gaussian. In accordance with the literature (see Bragg\cite{bragg}), we use a mean value
for the radius of $\mathbb{E}(R) = 1.39 \%$ of the chord length, and a mean
value for the position of $\mathbb{E}(S_r) = 20 \%$ of the chord length.

We select a profile for the mean horn geometry from Papadakis\cite{papadakis}. The two
independent stochastic parameters which govern perturbations from the mean shape
are the height $H$ (normal to the airfoil surface) and separation distance $S_h$
of the horn. The mean horn profile used is shown in
Fig.~\ref{fig:HornRidgeParam}.

\subsection{Ridge Ice Case}

\begin{figure}[htb]
 \begin{subfigmatrix}{2}
  \subfigure{\includegraphics{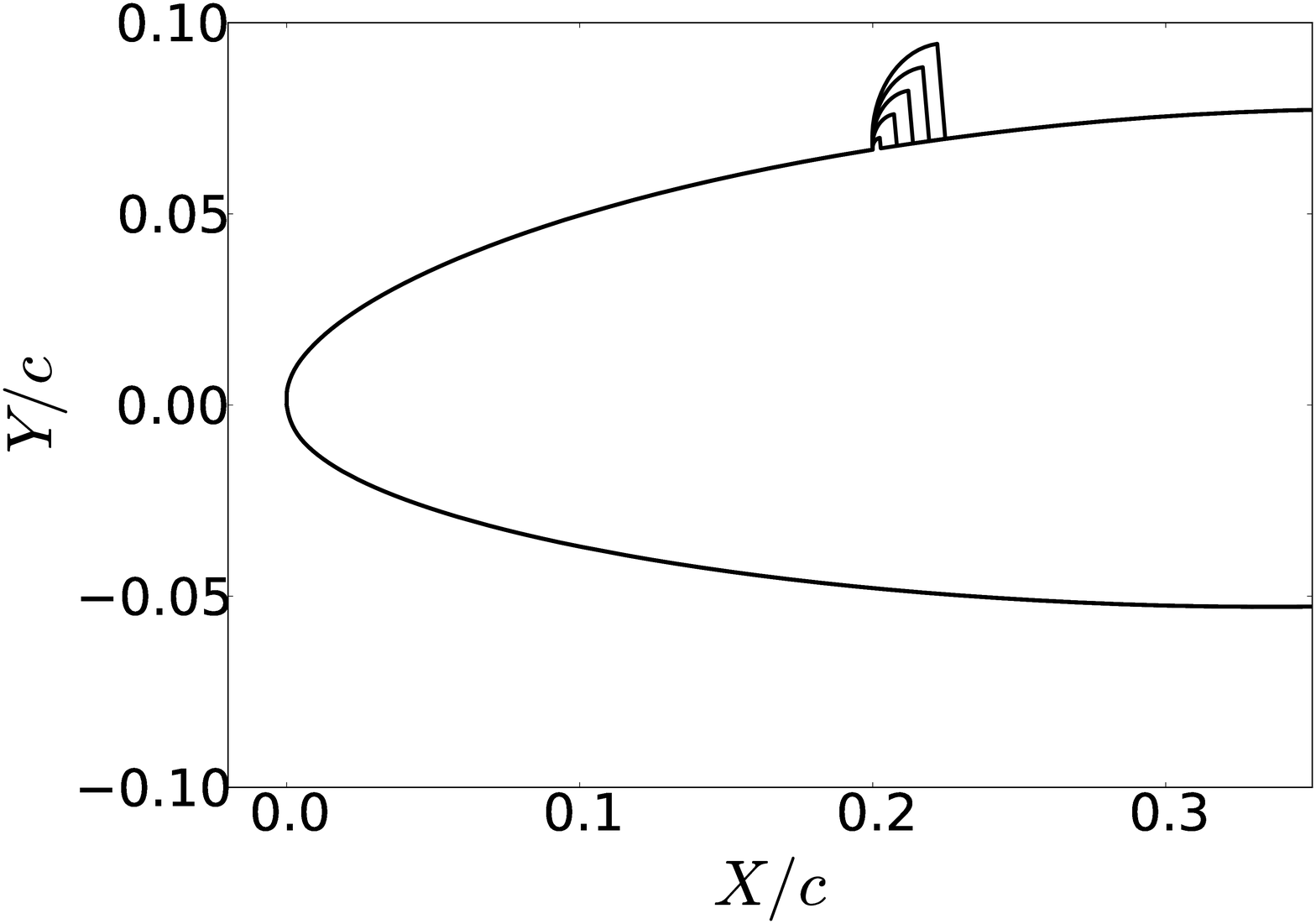}}
  \subfigure{\includegraphics{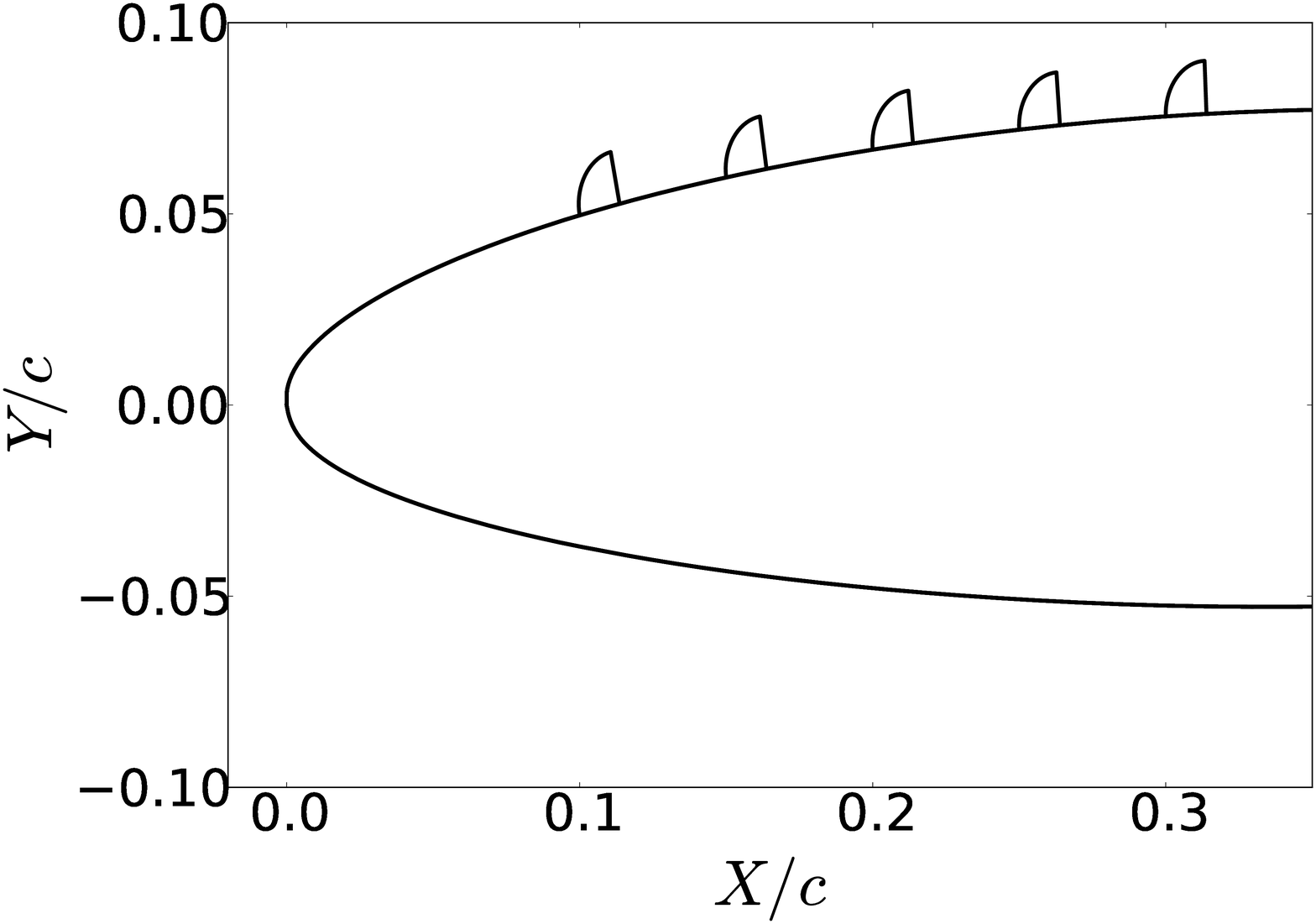}}
 \end{subfigmatrix}
 \caption{{\it LEFT:} Ridge shapes produced by perturbing $R$ by $dR$ =
   \{-2, -1, 0, 1, 2\}$\sigma_R$ with $\sigma_R = 40\%$ ($S_r = \mu_{S_r}$). {\it RIGHT:} Ridge shapes produced by
   perturbing $S_r$ by $dS_r$ = \{-2, -1, 0, 1, 2\}$\sigma_{S_r}$ with $\sigma_{S_r}
   = 5\%$ ($R = \mu_R$).}
 \label{fig:ridgeshapes}
\end{figure}

We begin with the ridge ice case, and quantify uncertainty in three separate
aerodynamic performace metrics: $\CLmax$, $\alphamax$, and
$\LDmax$. In each of the cases, there are two uncertain input parameters:
ridge radius, $R$, and ridge position, $S_r$. Both of these parameters assume
independent Gaussian distributions, where $\mu_R = 1.39\%$ of the chord, and
$\mu_{S_r} = 20\%$ of the chord; these were selected in agreement with
Bragg\cite{bragg}. We specify $\sigma_R$ as a percentage of $\mu_R$ and $\sigma_{S_r}$ as
a percentage of the chord length. We present two UQ investigations in which
the standard deviation pairs $(\sigma_R,\sigma_{S_r})$ take the values
$(10\%,1.25\%)$ and $(40\%,5\%)$.

Note that the flow solver used for all of our ridge and horn geometries is
FLO103, a 2D, steady-state Reynolds-Averaged Navier Stokes (RANS) solver
developed by Martinelli, Tatsumi, and Jameson \cite{martinelli}. It uses a
Spalart-Allmaras turbulence closure model, and uses no-injection, no-slip,
adiabatic, isothermal boundary conditions at the wall of the airfoil. Time
integration is performed explicitly by multigrid time stepping.

As mentioned, we compare our PCE methods to Quasi Monte Carlo
simulations. The basic algorithm used in this scheme involves inverse
transform sampling (see Devroye \cite{devroye}) for selecting samples. The
cumulative distribution space is sampled using an ergodic dynamical
system. This algorithm has a notable advantage of efficiency over
standard Monte Carlo algorithms: it can be proved that the samples
obtained through this Quasi Monte Carlo method converge in
distribution at a rate proportional to $1/N$, where $N$ is the number
of samples \cite{mezic}. This is in contrast to standard Monte Carlo,
in which the rate of convergence is $1/\sqrt{N}$.

\begin{figure}[H]
 \begin{subfigmatrix}{3}
  \subfigure[$\CLmax (dR,dS_r)$]{\includegraphics{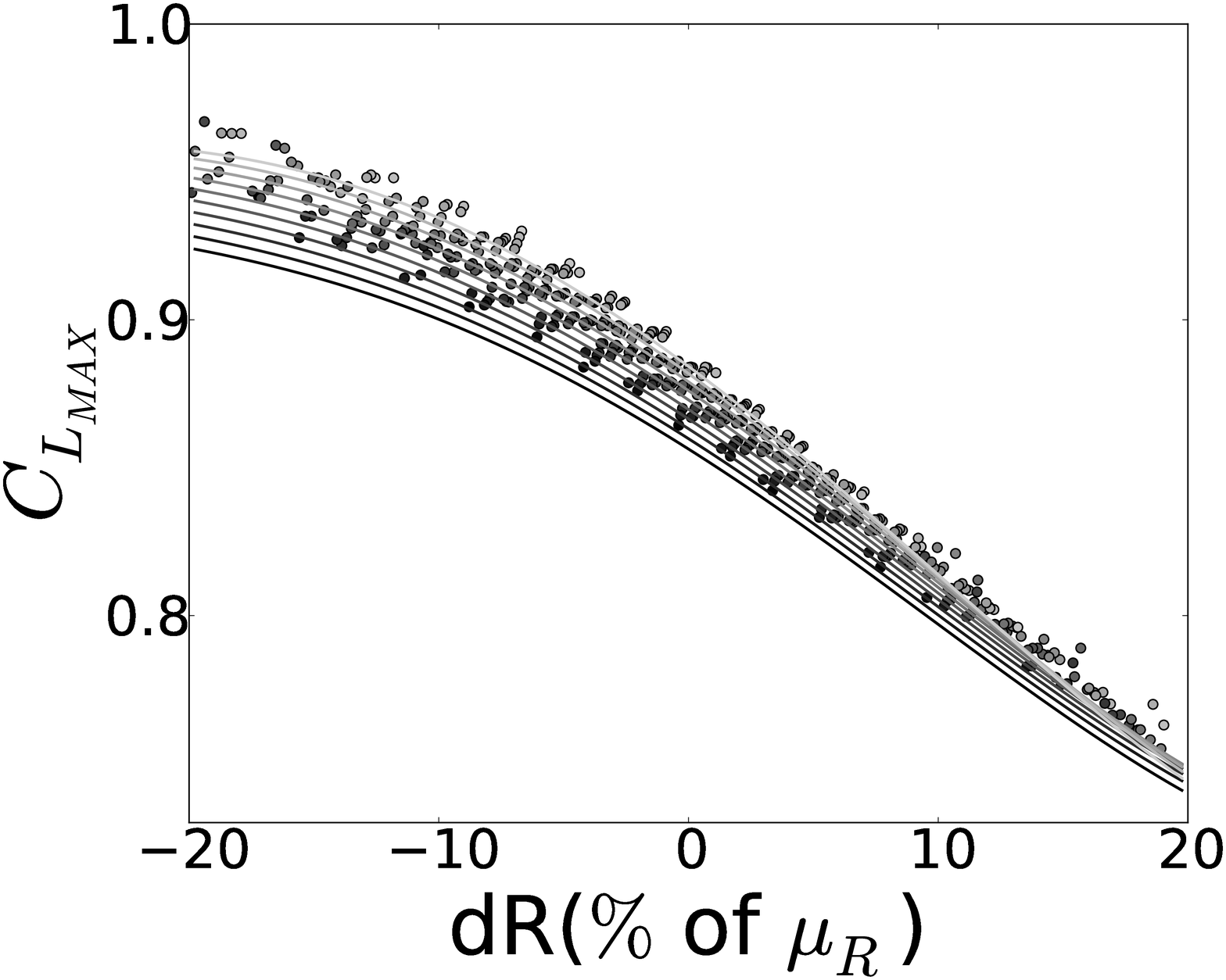}}
  \subfigure[$\alphamax (dR,dS_r)$]{\includegraphics{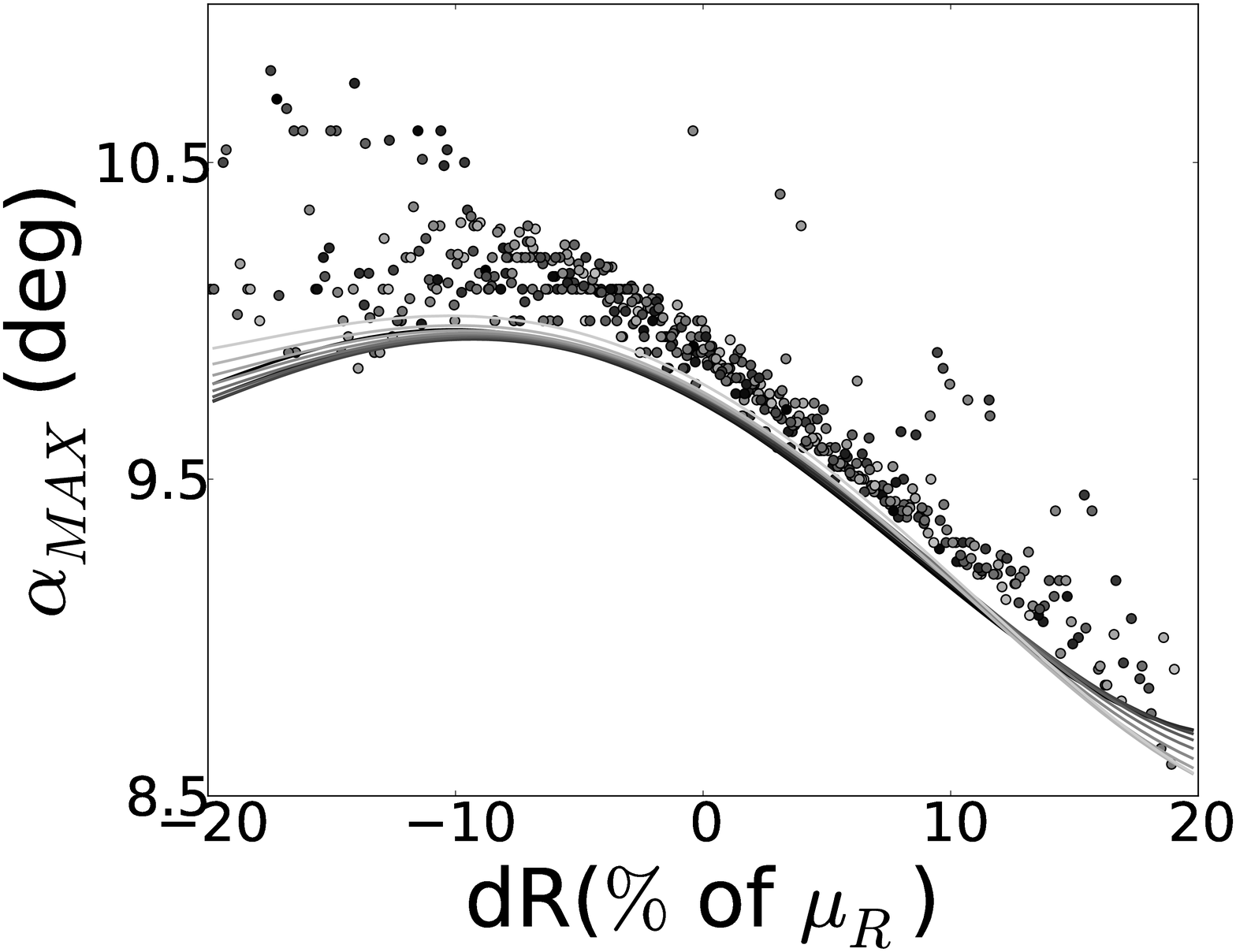}}
  \subfigure[$\LDmax (dR,dS_r)$]{\includegraphics{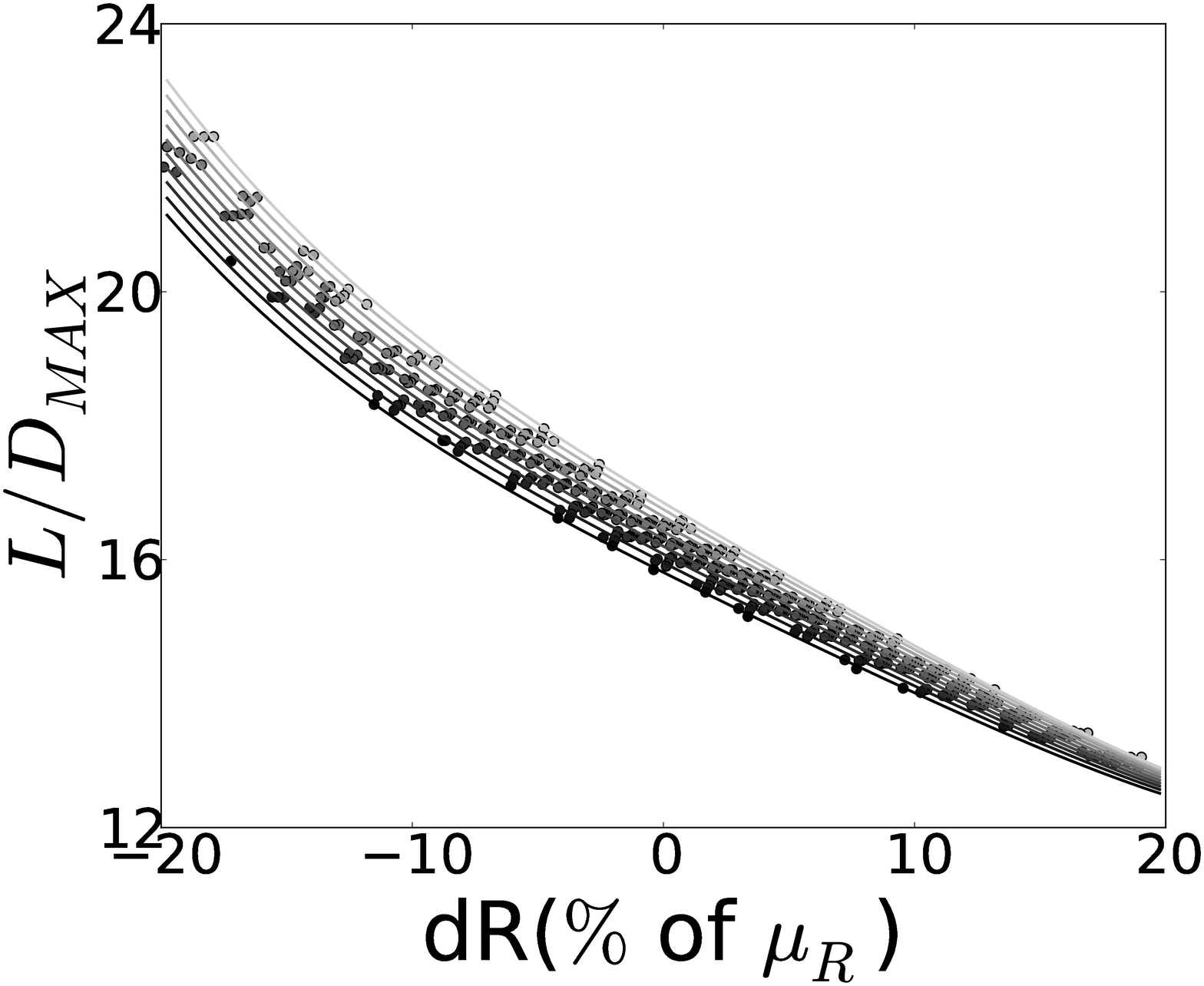}}
  \subfigure[PDF($\CLmax$)]{\includegraphics{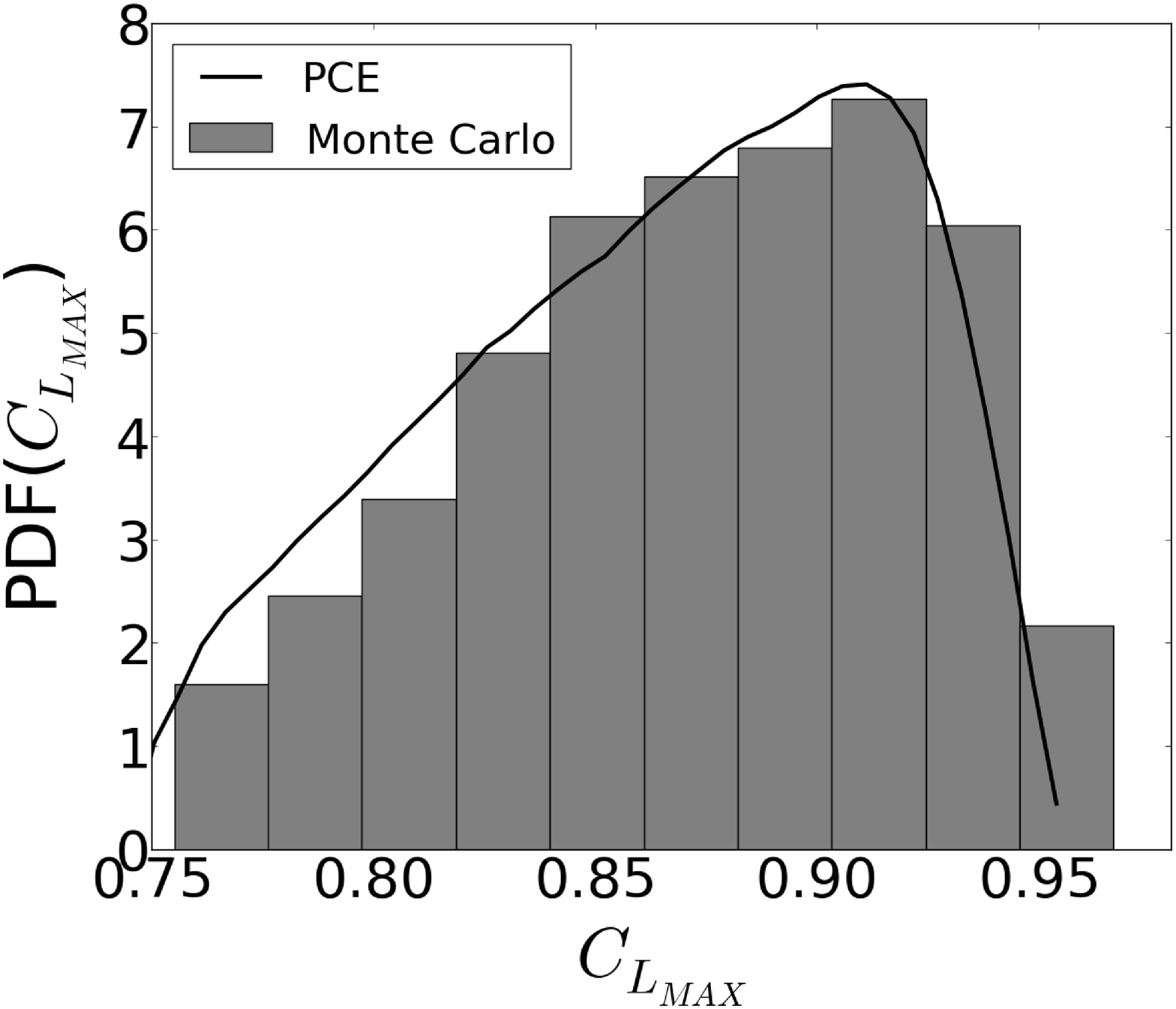}}
  \subfigure[PDF($\alphamax$)]{\includegraphics{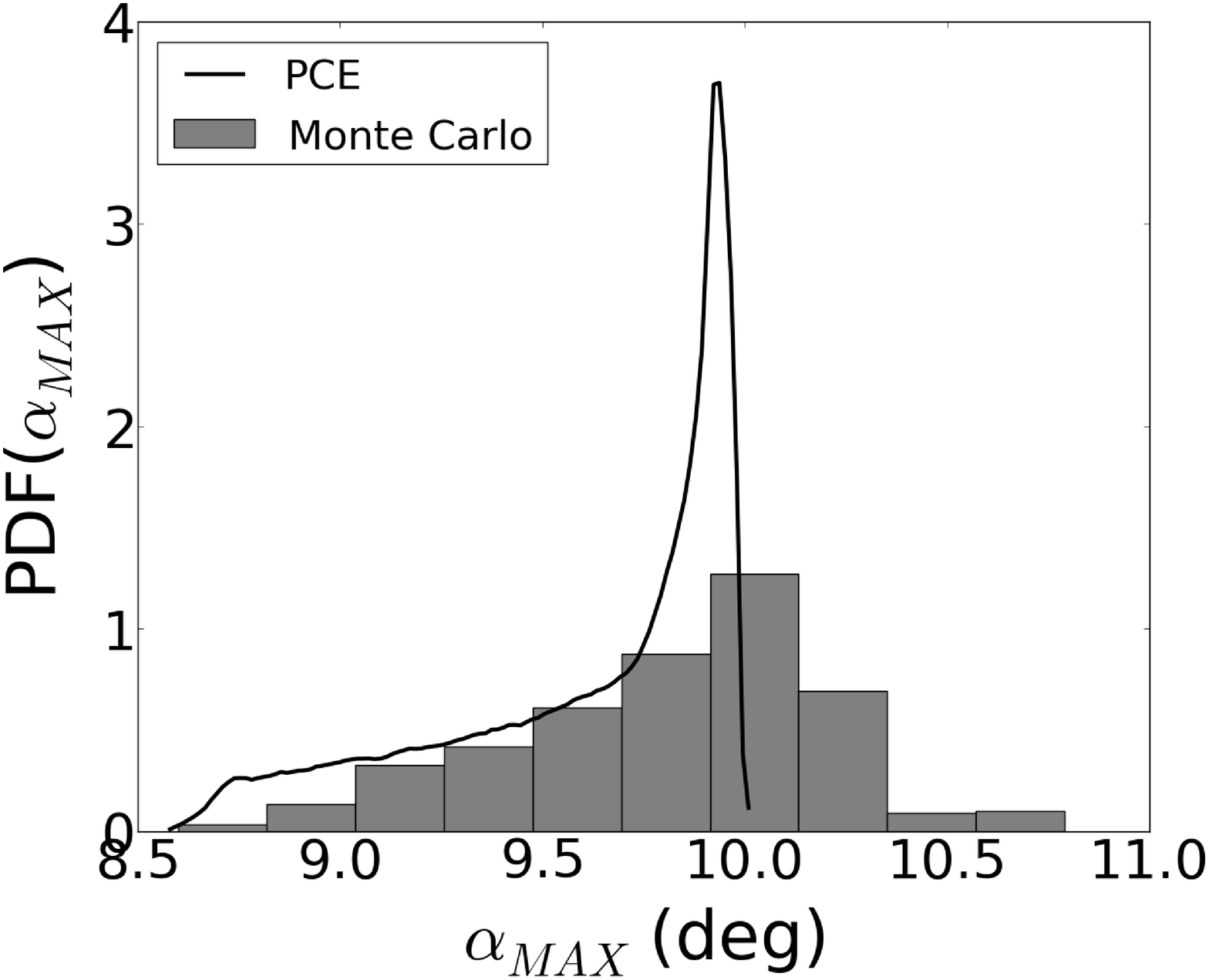}}
  \subfigure[PDF($\LDmax$)]{\includegraphics{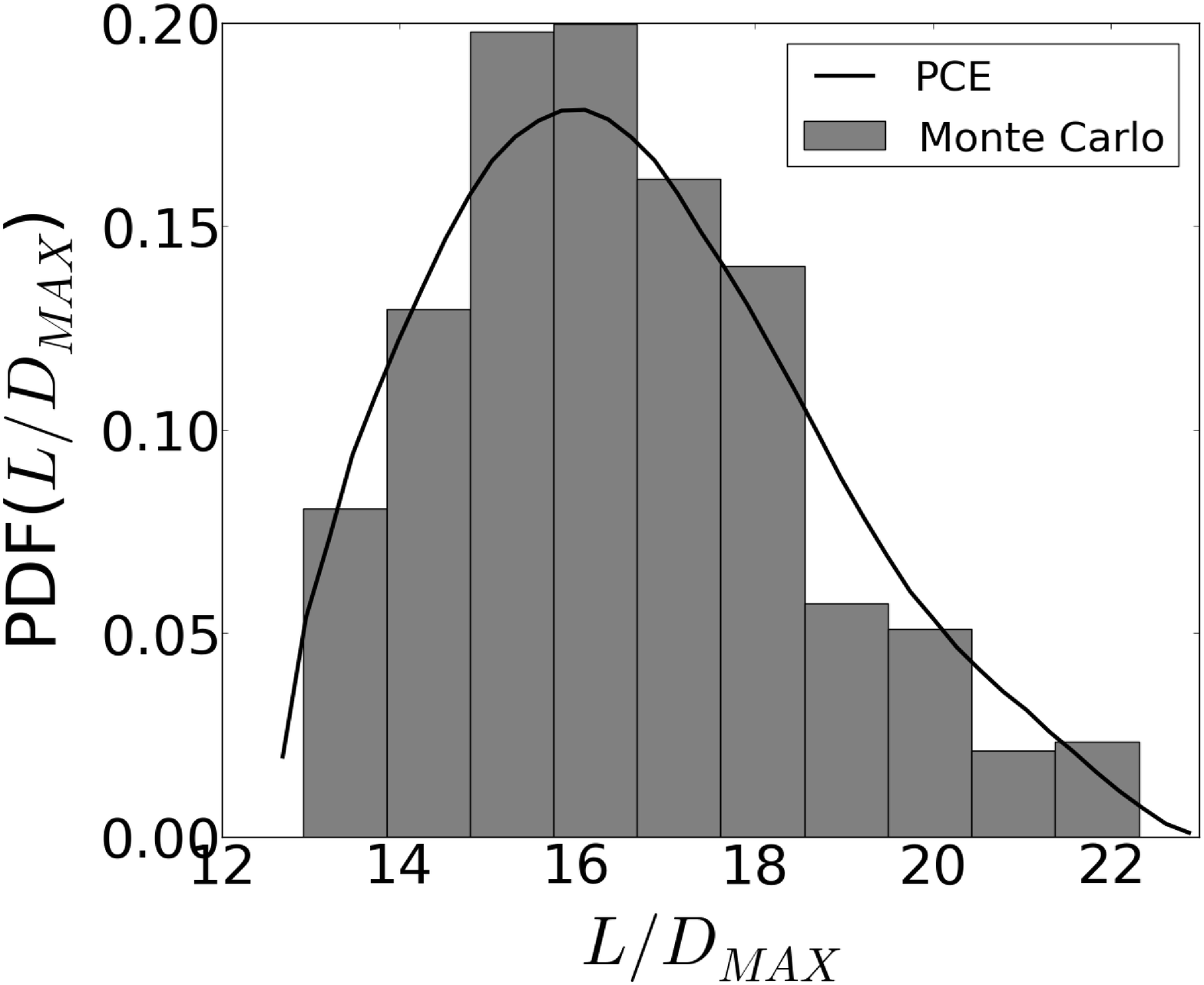}}
 \end{subfigmatrix}
 \caption{{\it TOP:} Comparisons of PCE surrogate maps to Quasi Monte Carlo
   results. Here $dR$ varies between $\pm 20\%$ of $\mu_R$, and
   $dS_r$ varies between $\pm 2.5\%$ of the chord length. The grayscale is chosen to
   represent different values of $dS_r$ (dark to light transition indicates
   increasing values of $dS_r$). The 10 PCE curves in each plot
   represent values of $dS_r$ equally-spaced between $\pm 2.5\%$.
{\it BOTTOM:} Comparisons of the normalized PDFs
   for both the Monte Carlo and PCE cases. The input distribution was a Gaussian
   with $\sigma_R$ = 10\% of $\mu_R$ and $\sigma_{S_r}$ = 1.25\% of the chord
   length, with both variables truncated at $2\sigma$.}
 \label{fig:surrogates_small}
\end{figure}

\newcommand{\ra}[1]{\renewcommand{\arraystretch}{#1}}
\newcommand{\squeezeup}{\vspace{-.5in}}
\begin{table*}\centering
  \caption{Comparison of Statistical Moments for Monte Carlo and PCE: ($\sigma_R,\sigma_{S_r}$) = (10\%,1.25\%)}
\ra{1}
\begin{tabular}{@{}rrrcrrcrr@{}}\toprule \toprule
& \multicolumn{2}{c}{$\CLmax$} & \phantom{abc}& \multicolumn{2}{c}{$\alphamax$ (deg)} &
\phantom{abc} & \multicolumn{2}{c}{$\LDmax$}\\
\cmidrule{2-3} \cmidrule{5-6} \cmidrule{8-9}
& MC & PCE && MC & PCE && MC & PCE \\ \midrule
Mean     & 0.87 & 0.86     && 9.8 & 9.6    && 16.6 & 16.5 \\
Variance & 0.0024 & 0.0025 && 0.16 & 0.13    && 4.0 & 4.4 \\
Skewness & $-0.33$ & $-0.37$   && $-0.50$ & $-0.96$    && 0.56 & 0.40 \\
Kurtosis & 2.3 & 2.2       && 2.9 & 2.7    && 3.0 & 2.6 \\
\bottomrule
\end{tabular}
\end{table*}

\begin{figure}[H]
 \begin{subfigmatrix}{3}
  \subfigure[$\CLmax (dR,dS_r)$]{\includegraphics{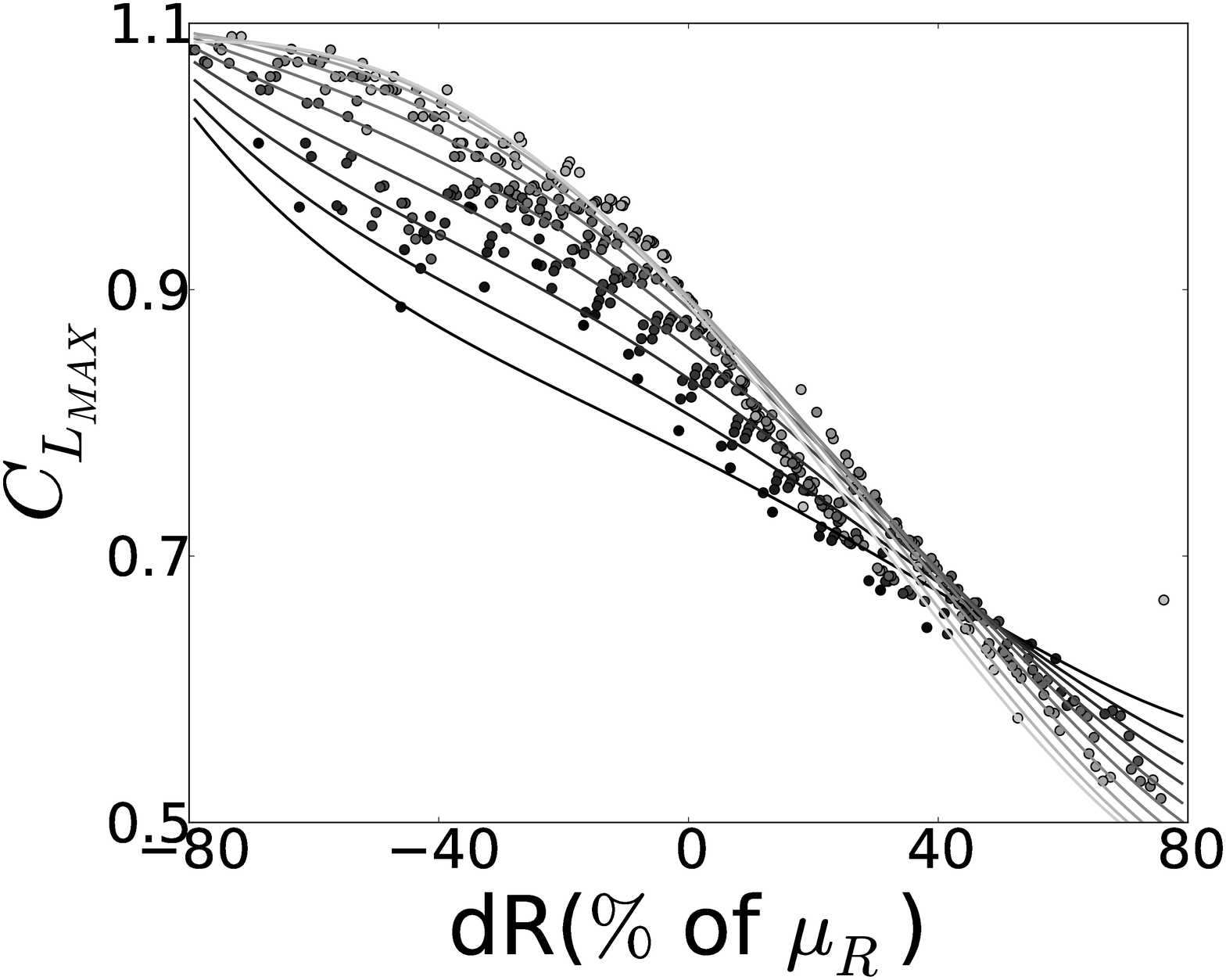}}
  \subfigure[$\alphamax (dR,dS_r)$]{\includegraphics{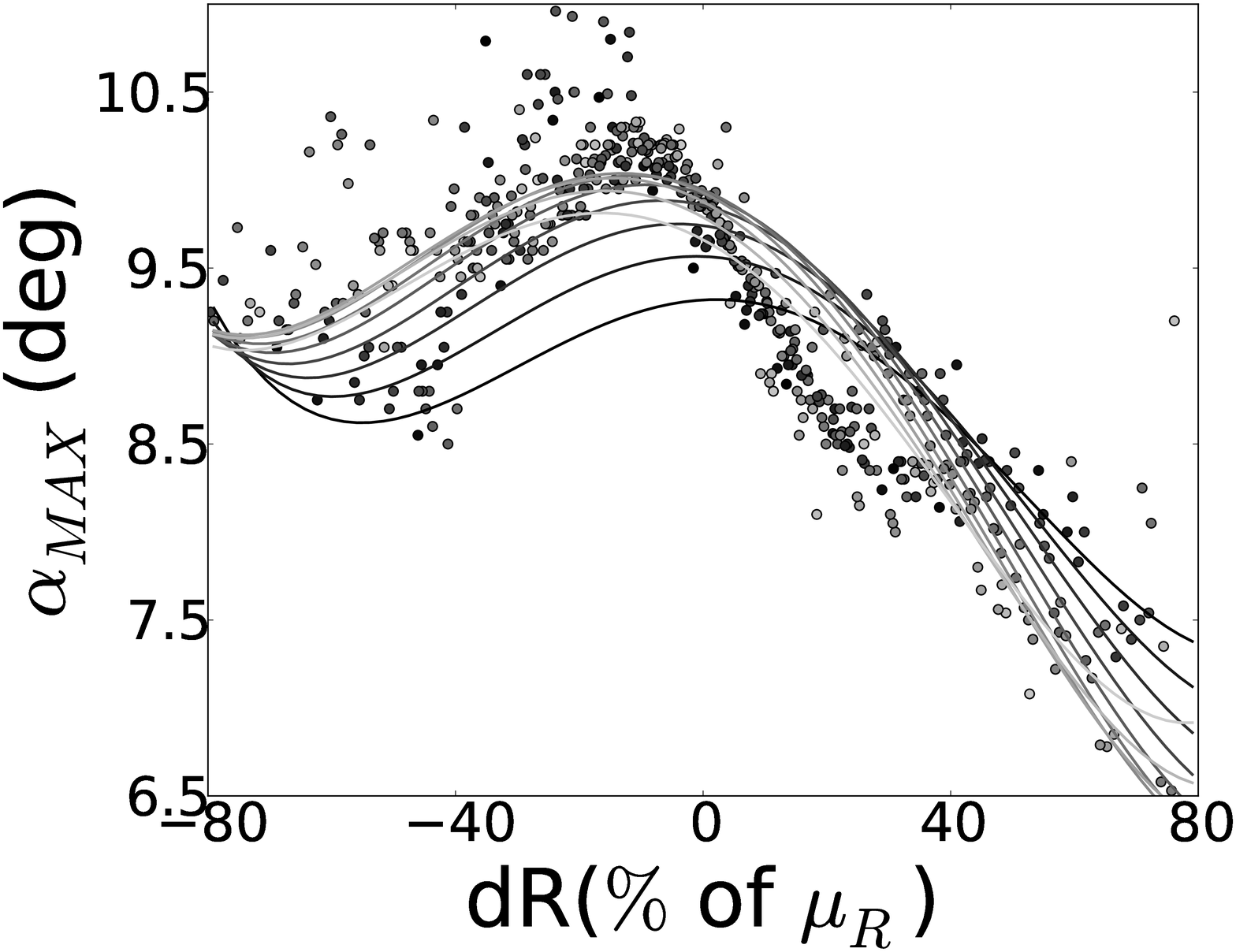}}
  \subfigure[$\LDmax (dR,dS_r)$]{\includegraphics{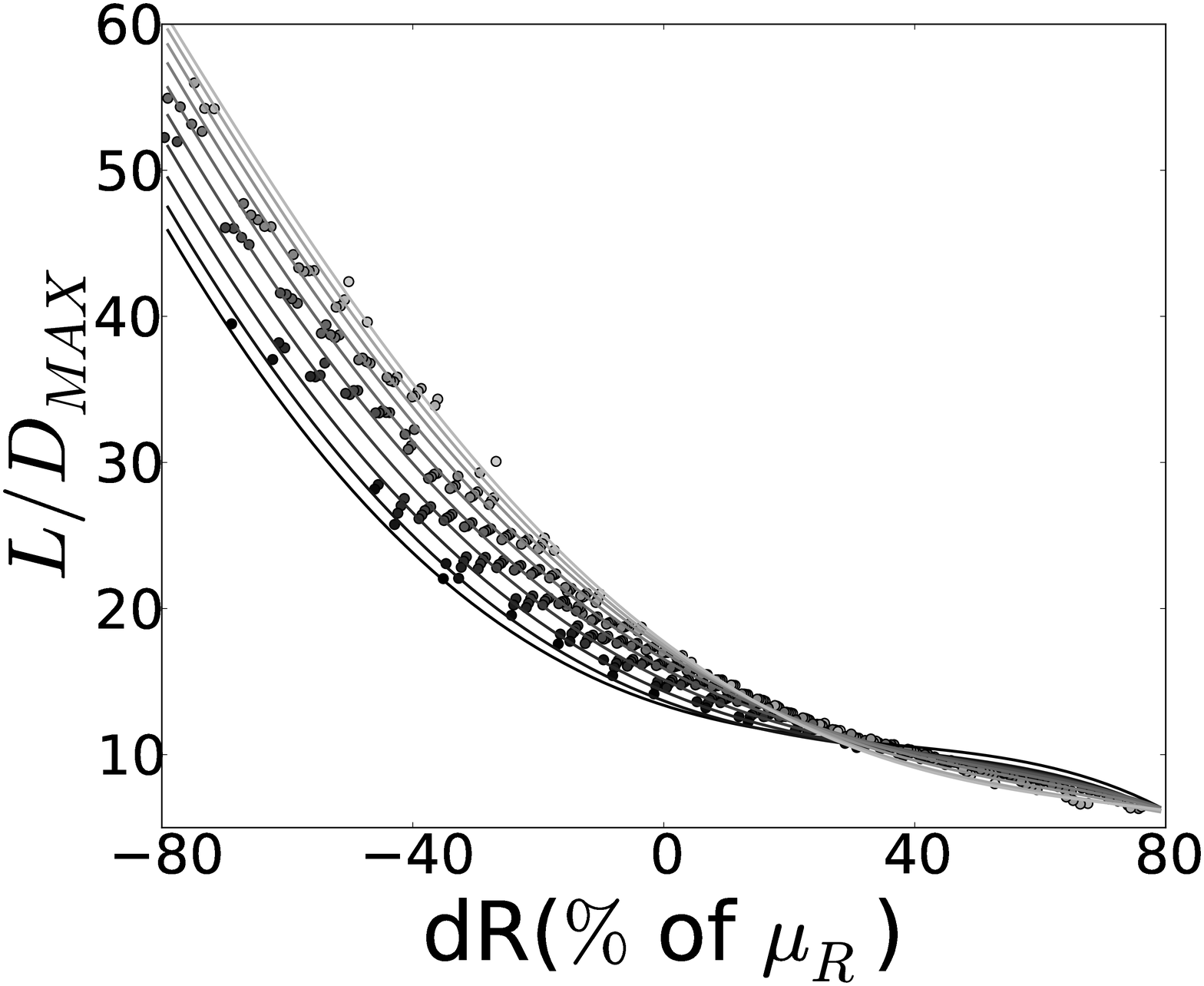}}
  \subfigure[PDF($\CLmax$)]{\includegraphics{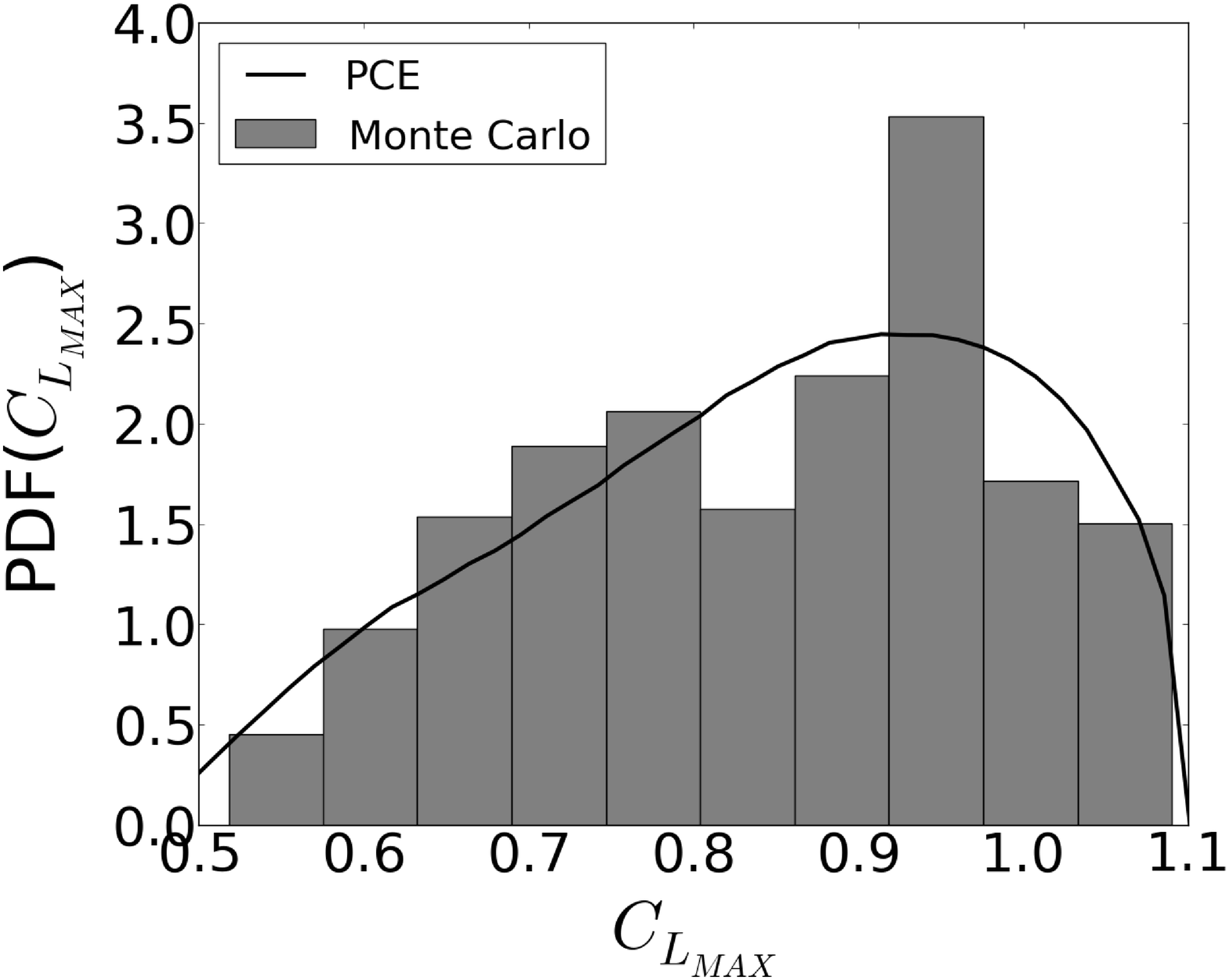}}
  \subfigure[PDF($\alphamax$)]{\includegraphics{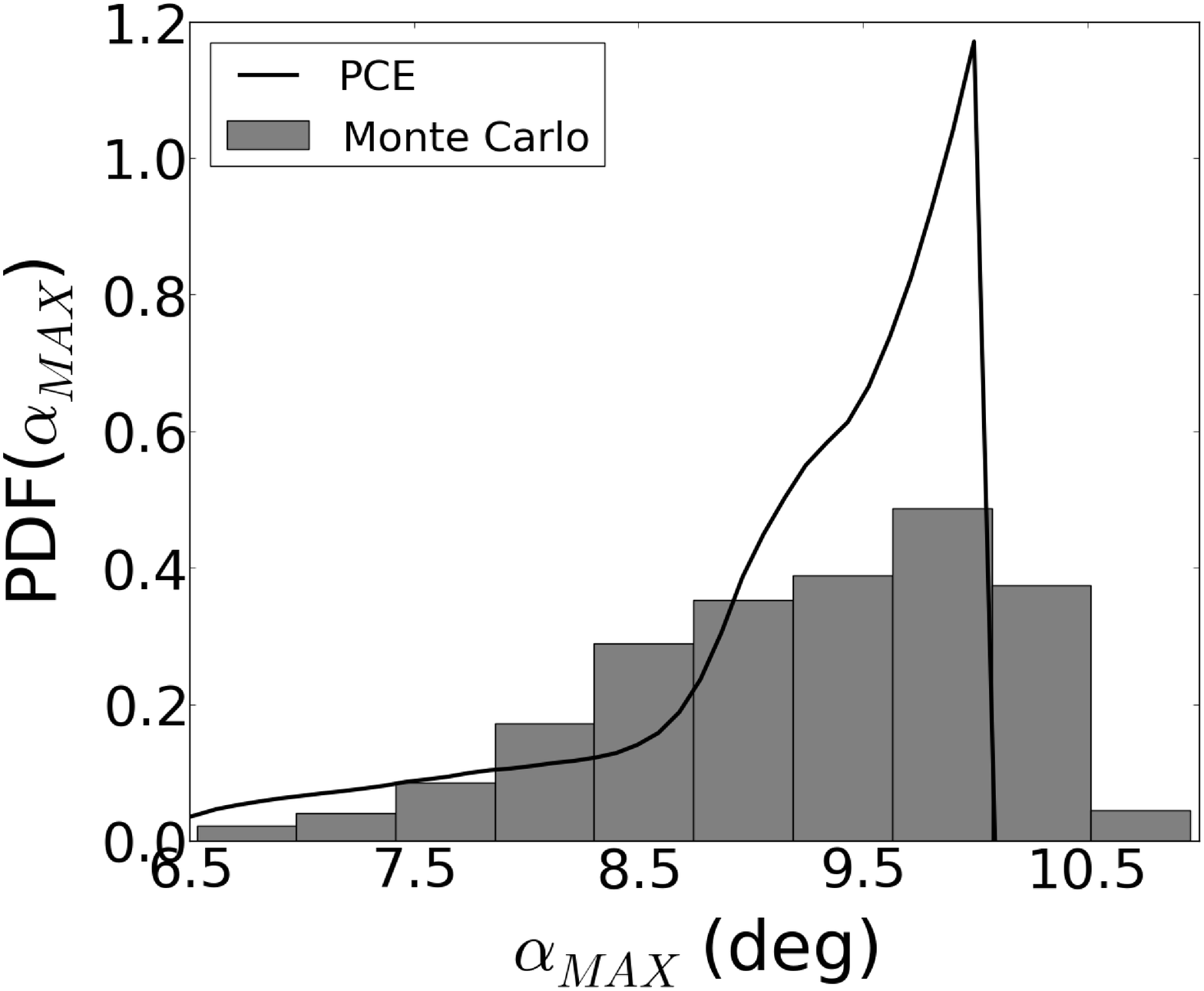}}
  \subfigure[PDF($\LDmax$)]{\includegraphics{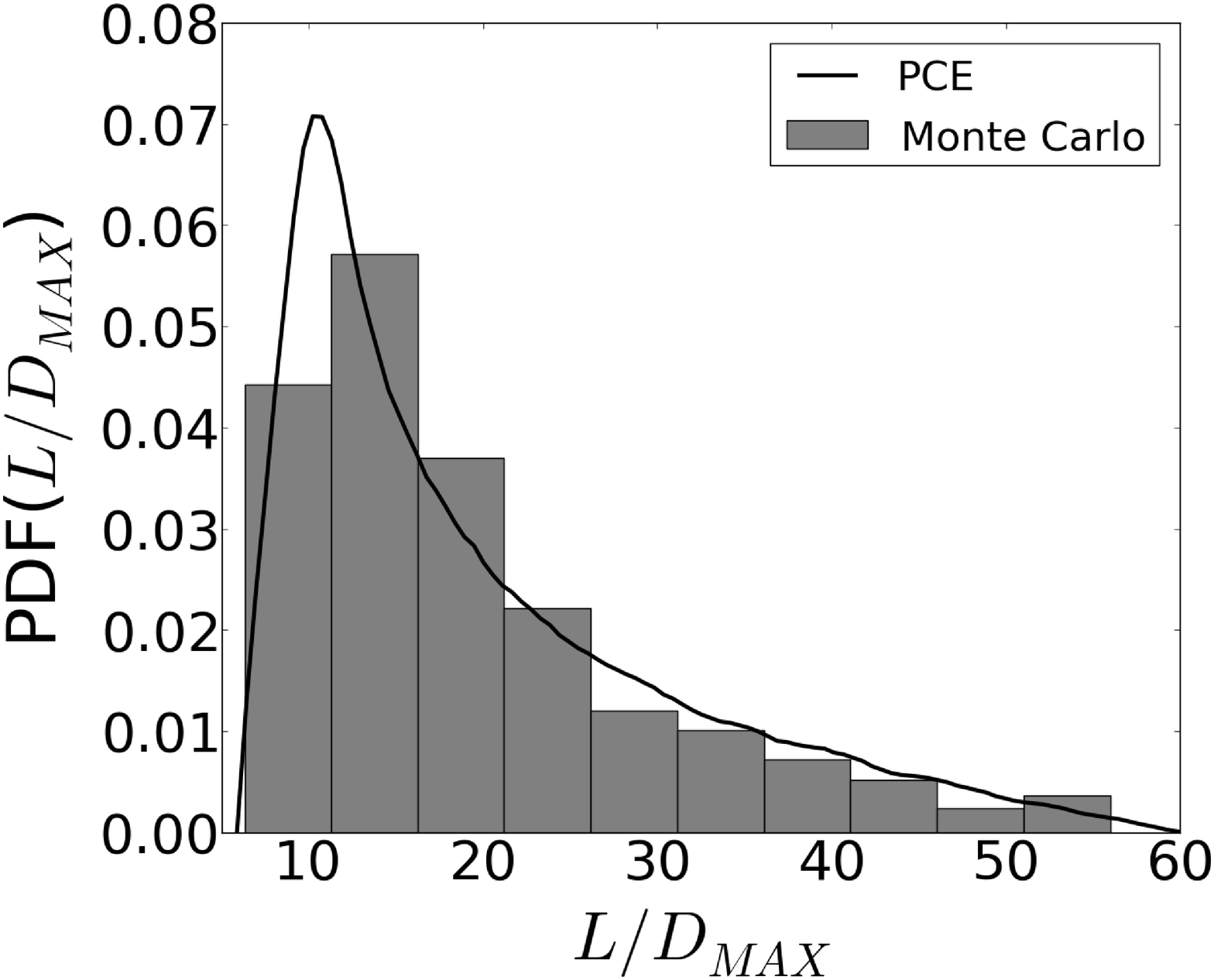}}
 \end{subfigmatrix}
 \caption{{\it TOP:} Comparisons of PCE surrogate maps to Quasi Monte Carlo
   results. Here $dR$ varies between $\pm 80\%$ of $\mu_R$, and
   $dS_r$ varies between $\pm 10\%$ of the chord length.
   The grayscale is chosen to
   represent different values of $dS_r$ (dark to light transition indicates
   increasing values of $dS_r$). The 10 PCE curves in each plot
   represent values of $dS_r$ equally-spaced between $\pm 10\%$ of the
   chord length.
{\it BOTTOM:} Comparisons of the normalized PDFs
   for both the Monte Carlo and PCE cases. The input distribution was a Gaussian
   with $\sigma_R$ = 40\% of $\mu_R$ and $\sigma_{S_r}$ = 5\% of the chord
   length, with both variables truncated at $2 \sigma$.}
 \label{fig:surrogates_large}
\end{figure}

\begin{table*}\centering
  \caption{Comparison of Statistical Moments for Monte Carlo and PCE: ($\sigma_R,\sigma_{S_r}$) = (40\%,5\%)}
\ra{1}
\begin{tabular}{@{}rrrcrrcrr@{}}\toprule \toprule
& \multicolumn{2}{c}{$\CLmax$} & \phantom{abc}& \multicolumn{2}{c}{$\alphamax$ (deg)} &
\phantom{abc} & \multicolumn{2}{c}{$\LDmax$}\\
\cmidrule{2-3} \cmidrule{5-6} \cmidrule{8-9}
& MC & PCE && MC & PCE && MC & PCE \\ \midrule
Mean     & 0.85 & 0.85     && 9.2 & 9.2      && 19.4 & 19.7 \\
Variance & 0.020 & 0.022   && 0.72 & 0.69    && 110 & 120 \\
Skewness & $-0.29$ & $-0.38$   && $-0.60$ & $-1.4$   && 1.3 & 1.2 \\
Kurtosis & 2.1 & 2.2       && 2.9 & 4.4      && 4.3 & 3.7 \\
\bottomrule
\end{tabular}
\end{table*}

The top rows of Fig.~\ref{fig:surrogates_small}--\ref{fig:surrogates_large}
compare the surrogate maps created using the PCE stochastic collocation method
to results obtained through Quasi Monte Carlo sampling for $\CLmax$,
$\alphamax$, and $\LDmax$. The bottom rows present normalized histograms
of the Quasi Monte Carlo results and compare them to the results of propagating
the input distribution through the PCE surrogate map. The Quasi Monte Carlo
method used 500 samples per case, whereas the PCE method utilized 4$^{th}$ order
polynomial expansions and hence required a 5$\times$5 collocation
mesh. It should be noted that the Jacobi polynomials---not the Hermite
polynomials---were used as the basis in the PCE scheme. There are two
reasons for this choice. First, the input distribution in all cases
was a truncated Gaussian (i.e., both $dR$ and $dS_r$ were truncated at
$\pm 2\sigma$). Second, use of the Hermite polynomials might have
presented a practical problem, as the collocation nodes tend to lay
far out in the tails of the distribution. Sampling at these extreme
nodes (corresponding to extreme ridge radii and/or positions) might
have presented a problem for the RANS solver. Hence, we used a Jacobi
expansion which approximates a truncated Gaussian in distribution and
does not require sampling at extreme positions. Specifically, denoting
the univariate Jacobi polynomials as $\{ J_i^{(\alpha,\beta)}\}$, we used the linear expansion $\sum_{i=0}^5 a_iJ_i^{(2,2)}$ to approximate a truncated Gaussian. See
Xiu\cite{xiu_book} for an in-depth discussion and for the numerical
values of the coefficients $a_i$.

Examining the data, we observe that the dominant parameter in all cases is the ridge radius, $R$; the ridge position $S_r$ has a
relatively smaller effect on the metrics. A larger ridge size (i.e., increasing
$dR$) leads to a monotonic decrease in $\CLmax$ and $\LDmax$. A ridge
which is closer to the leading edge (i.e., decreasing $dS_r$) also results in a
monotonic decrease in $\CLmax$ and $\LDmax$. This is quite intuitive,
since a large ridge radius tends to promote large scale flow separation at lower
angles of attack, and a ridge closer to the leading edge disrupts more of the
flow over the airfoil.

It is also clear that the agreement between the PCE and Monte Carlo schemes is
best for the metrics $\CLmax$ and $\LDmax$; agreement for $\alphamax$
is less satisfactory. This can be attributed to the smoothness of the maps. The
maps $\CLmax(dR,dS_r)$ and $\LDmax(dR,dS_r)$ are both very smooth---in
spectral terms, most of the energy of the PCE expansions which approximate them
is contained in the low order modes.

In contrast, $\alphamax(dR,dS_r)$ is not as smooth, as can be seen in the Monte
Carlo results, particularly at extreme values of $dR$. This reflects a
nontrivial amount of energy in higher order spectral terms which are neglected
in our 4$^{th}$ order PCE expansions. However, this should be considered in
context with how we obtain the values of $\alphamax$. Our algorithm for
detecting $\CLmax$ and $\alphamax$ involves testing the curvature of
discrete points on the lift curve---if the curvature exceeds some calibrated
bounds, then stall is assumed to have occured, and $\CLmax$ and
$\alphamax$ are interpolated from the discrete points. As one might expect,
this method works well for detecting $\CLmax$, since that quantity is
relatively constant near stall. However, it does not perform as robustly for
$\alphamax$, since that quantity is much more sensitive to perturbations from
the true value near stall. Thus, in a sense, we do not even wish to recover the
higher order modes for $\alphamax$, since these reflect the lack of
robustness of our algorithm instead of the general trends in the parameter
space.

It should also be noted that if greater refinement of the PCE results for
$\alphamax$ is desired, this could be achieved either through retaining
higher order terms in the PCE expansion of $\alphamax$ ($p$-refinement), or
through dividing the stochastic space into smaller, discrete elements
($h$-refinement), or a combination of these tactics. In the next section on horn
icing, we explicitly demonstrate how to apply stochastic $h$-refinement to
improve PCE results.

\begin{figure}[htb]
 \begin{subfigmatrix}{3}
   \subfigure{\includegraphics{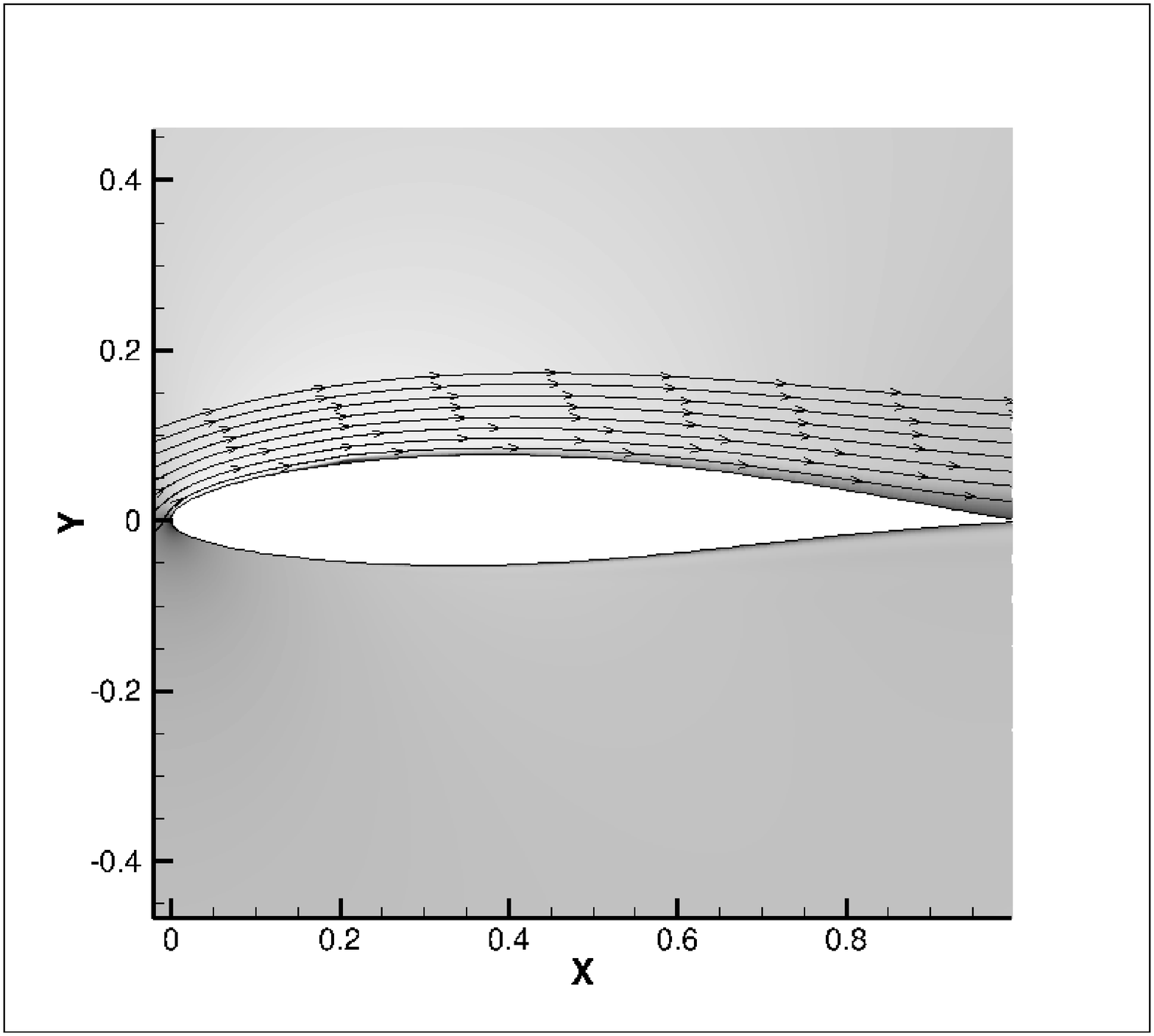}}
   \subfigure{\includegraphics{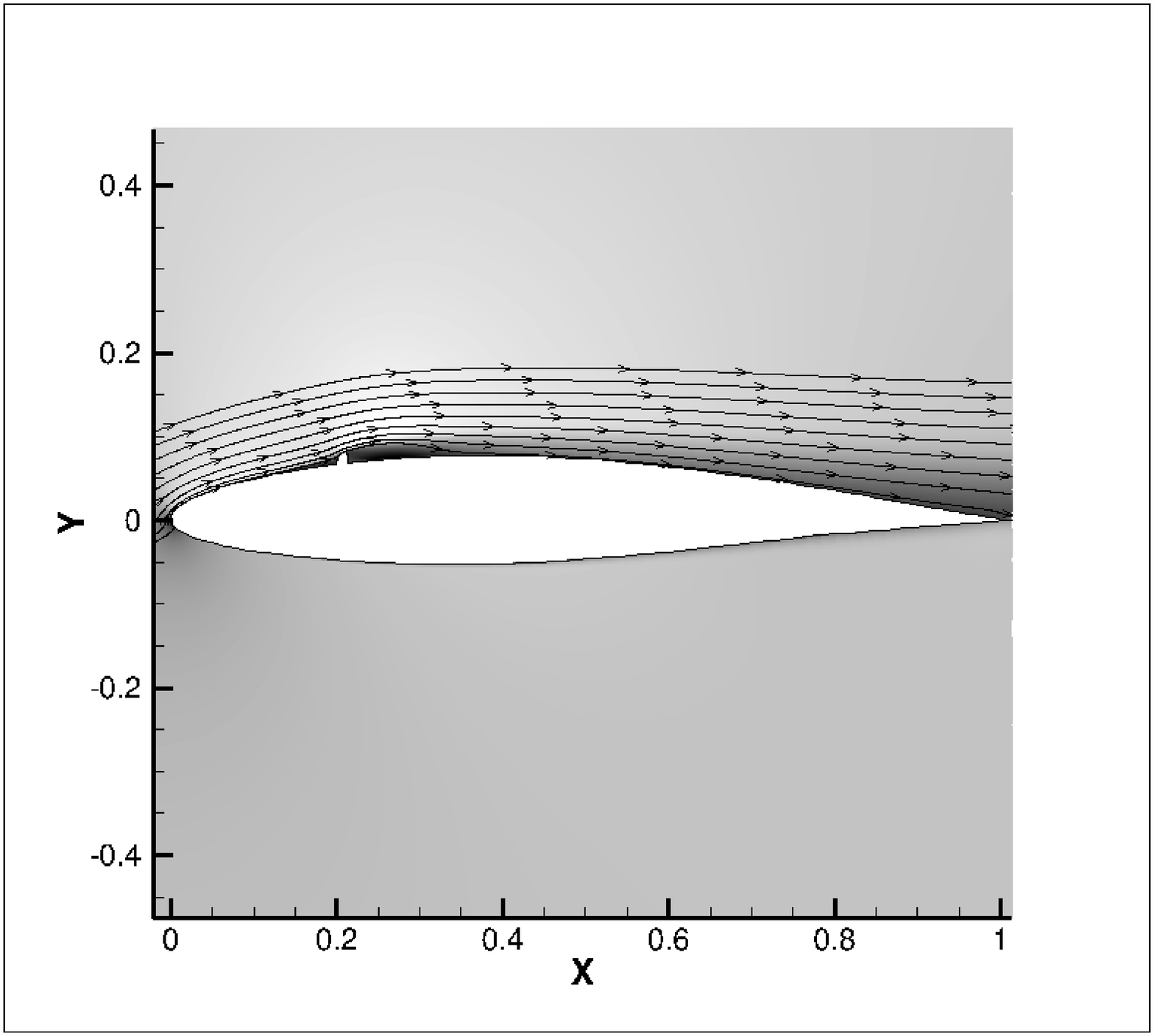}}
   \subfigure{\includegraphics{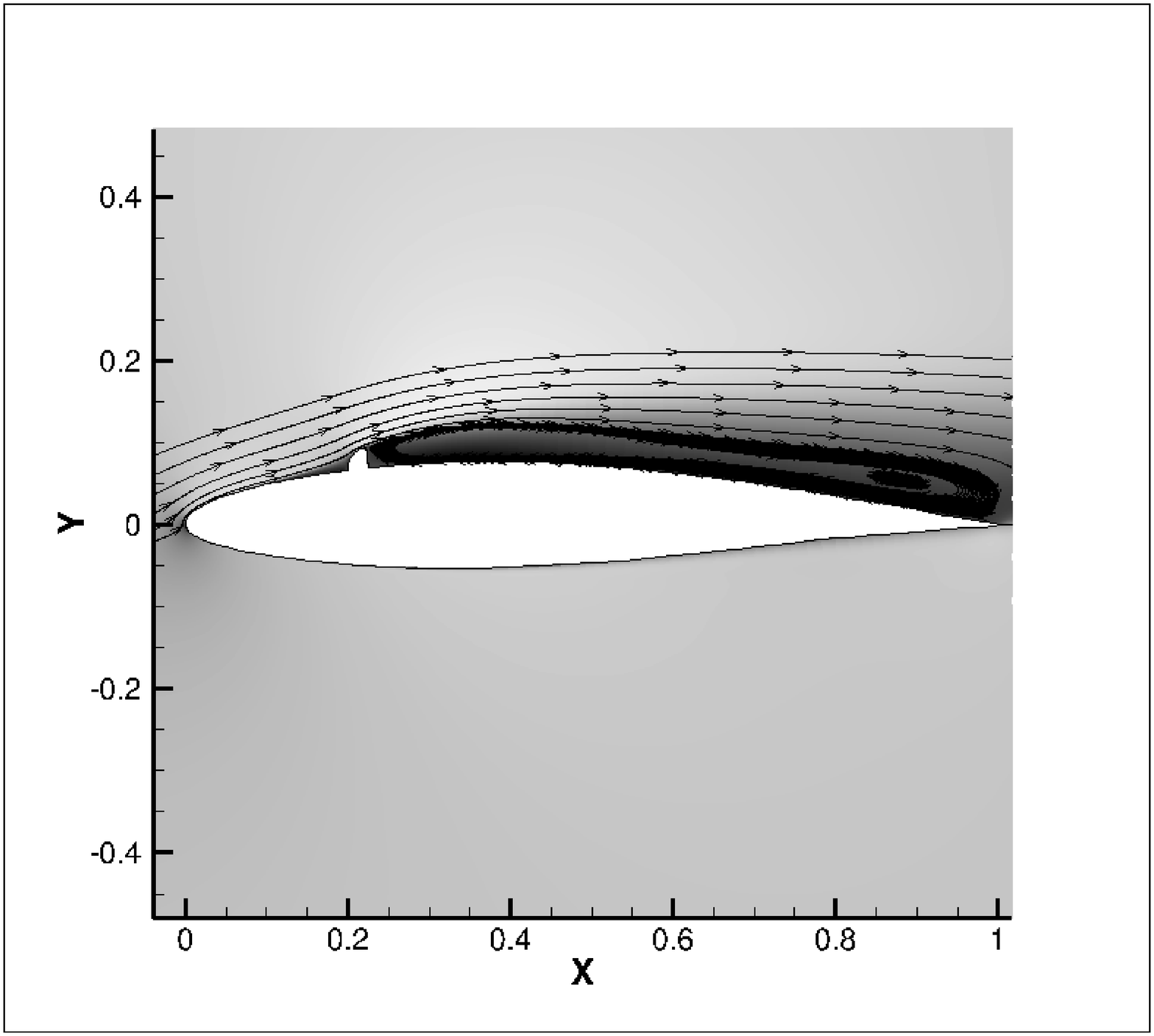}}
 \end{subfigmatrix}
 \caption{Flow field ($\rho u$) contours for the ridge ice case with increasing
   ridge radius at $\alpha = 6^\circ$ in which $dR = -80$, $0$, and $80 \%$ of
   $\mu_R$ (going from left to right). The solution for the smallest ridge
   closely resembles that of the clean airfoil. A large scale separation bubble
   forms aft of the ridge for the medium sized case before reattaching to the
   airfoil surface; this separation bubble reduces lift and increases drag. For
   the large size ridge, the separation bubble caused by the ridge is so large
   that the flow does not reattach, leading to early stall.}
 \label{fig:separation_ridge}
\end{figure}

The effect of an increasing ridge radius on the flowfield is shown in
Fig.~\ref{fig:separation_ridge}. This figure makes it very clear how different
our UQ study is from a sensitivity study, in which the effect of only small
perturbations is investigated. In contrast, we are investigating a large amount
of uncertainty in ridge size and position, and this leads to a large spectrum of
possible flowfields.

\subsection{Horn Ice Case}

We approach the horn ice UQ problem in the following way. First, we identify a
canonical horn ice shape, which was originally produced by NASA's LEWICE icing
code for 22.5 minutes of ice accretion on a NACA 63A213 airfoil (see
Papadakis\cite{papadakis} for more details). That shape is identified in
Fig.~\ref{fig:HornRidgeParam}

In our UQ framework, we allow a scaling of the horn height by a parameter which
we denote as $h$, with $h \in$ [0,1] (0 corresponds to the clean airfoil, and 1
corresponds to the horn height given in the figure). We also allow a scaling of
the inter-horn separation distance by a parameter which we denote as $s$, with
$s \in$ [0.1,1.9] (that is, the inter-horn separation distance in the figure can
vary by $\pm 90\%$ of its nominal value). We wish to investigate uncertainty in
aerodynamic performance metrics given uncertainty in $h$ and $s$, where $h$ is
the positive half of the Gaussian $\mathcal{N}(0,0.5^2)$ (which we will denote
as $\mathcal{N}^+_{1/2}(0,0.5^2)$), and $s$ is the Gaussian $\mathcal{N}(1,
0.45^2)$. It should be noted that both parameters are truncated at 2$\sigma$
distance from the mean.

\begin{figure}[htb]
 \begin{subfigmatrix}{2}
  \subfigure{\includegraphics{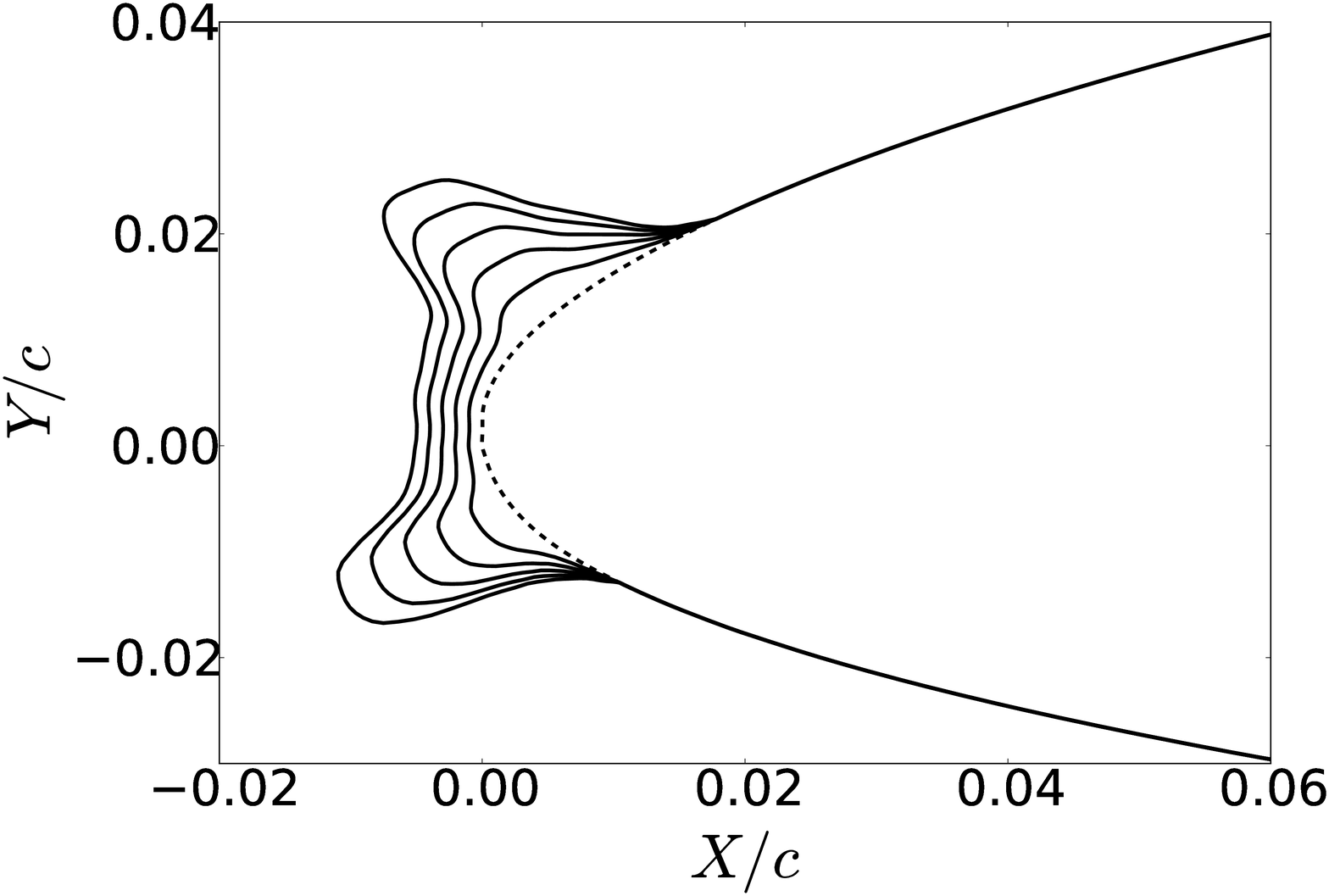}}
  \subfigure{\includegraphics{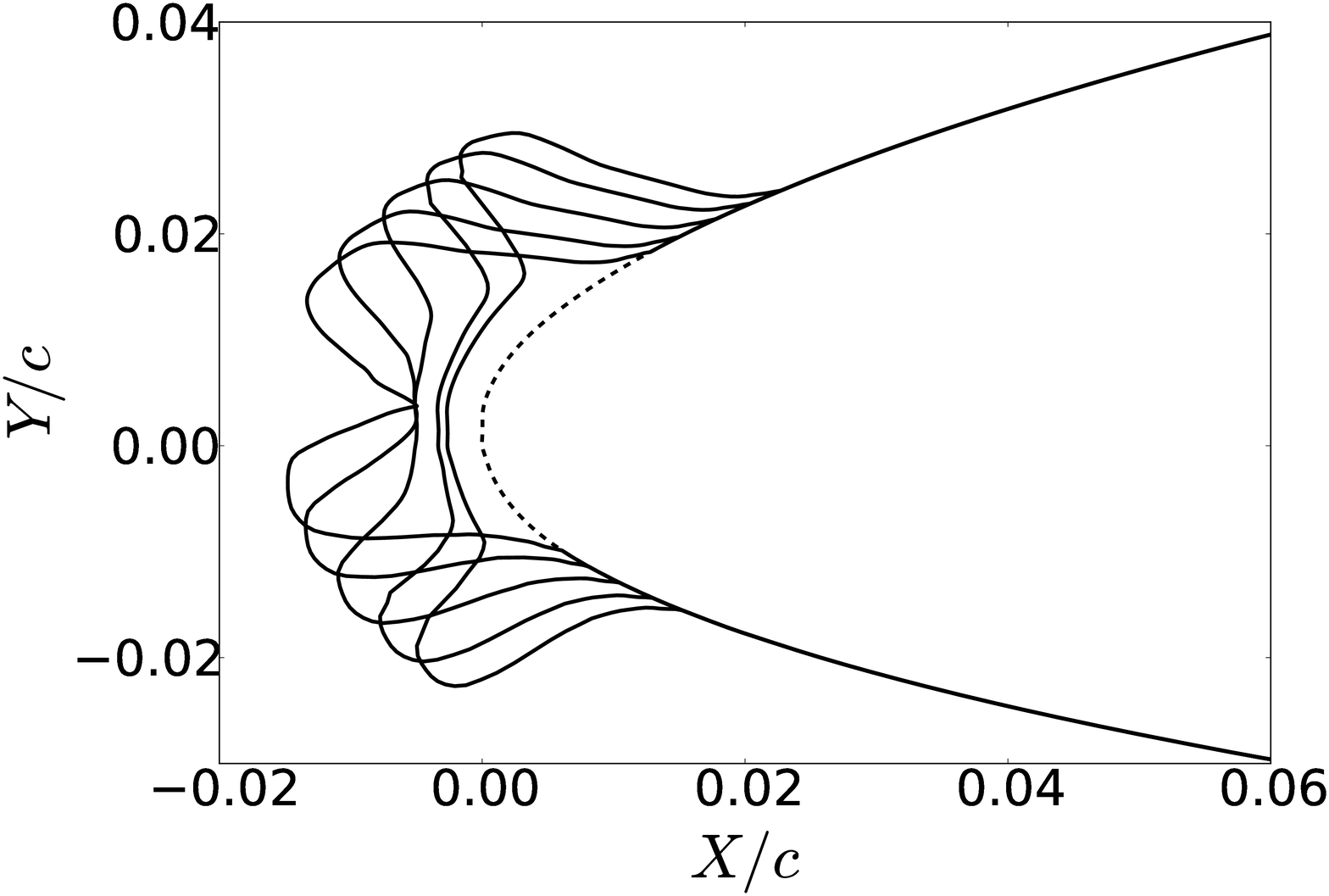}}
 \end{subfigmatrix}
 \caption{{\it LEFT:} Horn shapes produced by variation of the parameter $h\in
   \{0.2, 0.4, 0.6, 0.8, 1\}$ ($s = 1$). {\it RIGHT:} Horn shapes produced by
   variation of the parameter $s \in \{0.1, 0.5, 1, 1.5, 1.9\}$ ($h = 1$).}
 \label{fig:hornshapes}
\end{figure}

\begin{figure}[H]
 \begin{subfigmatrix}{3}
  \subfigure[$\CLmax (h,s)$]{\includegraphics{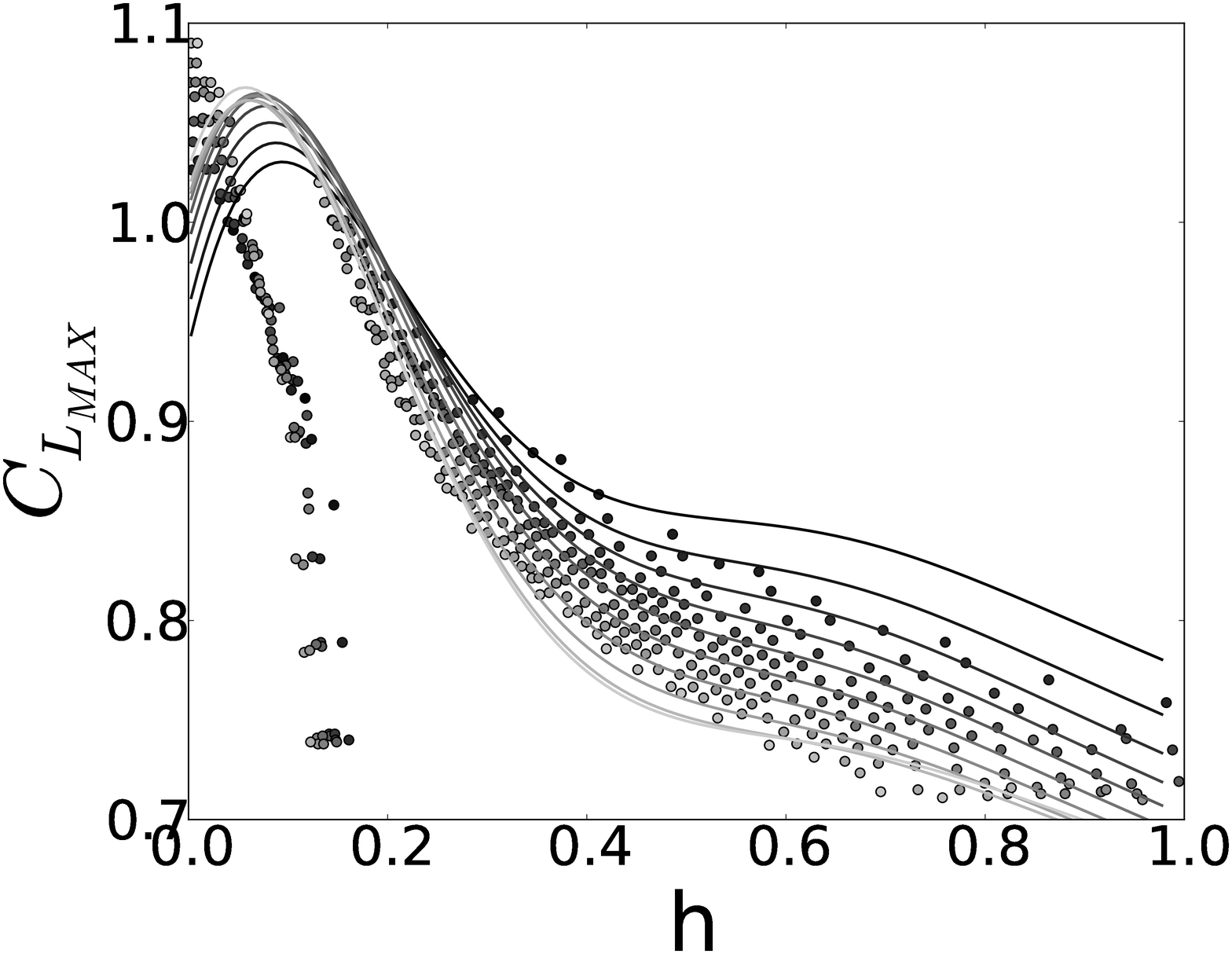}}
  \subfigure[$\alphamax (h,s)$]{\includegraphics{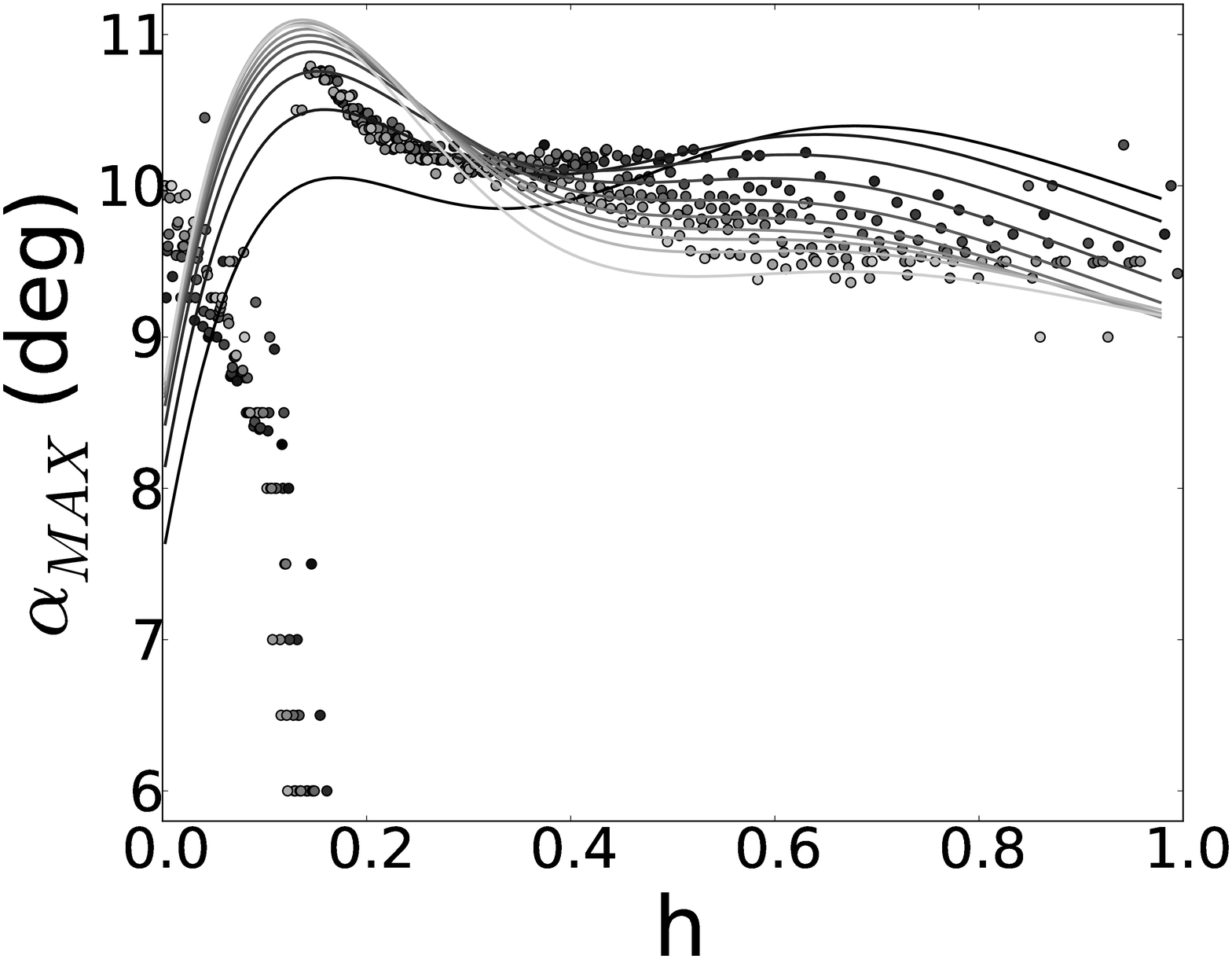}}
  \subfigure[$\LDmax (h,s)$]{\includegraphics{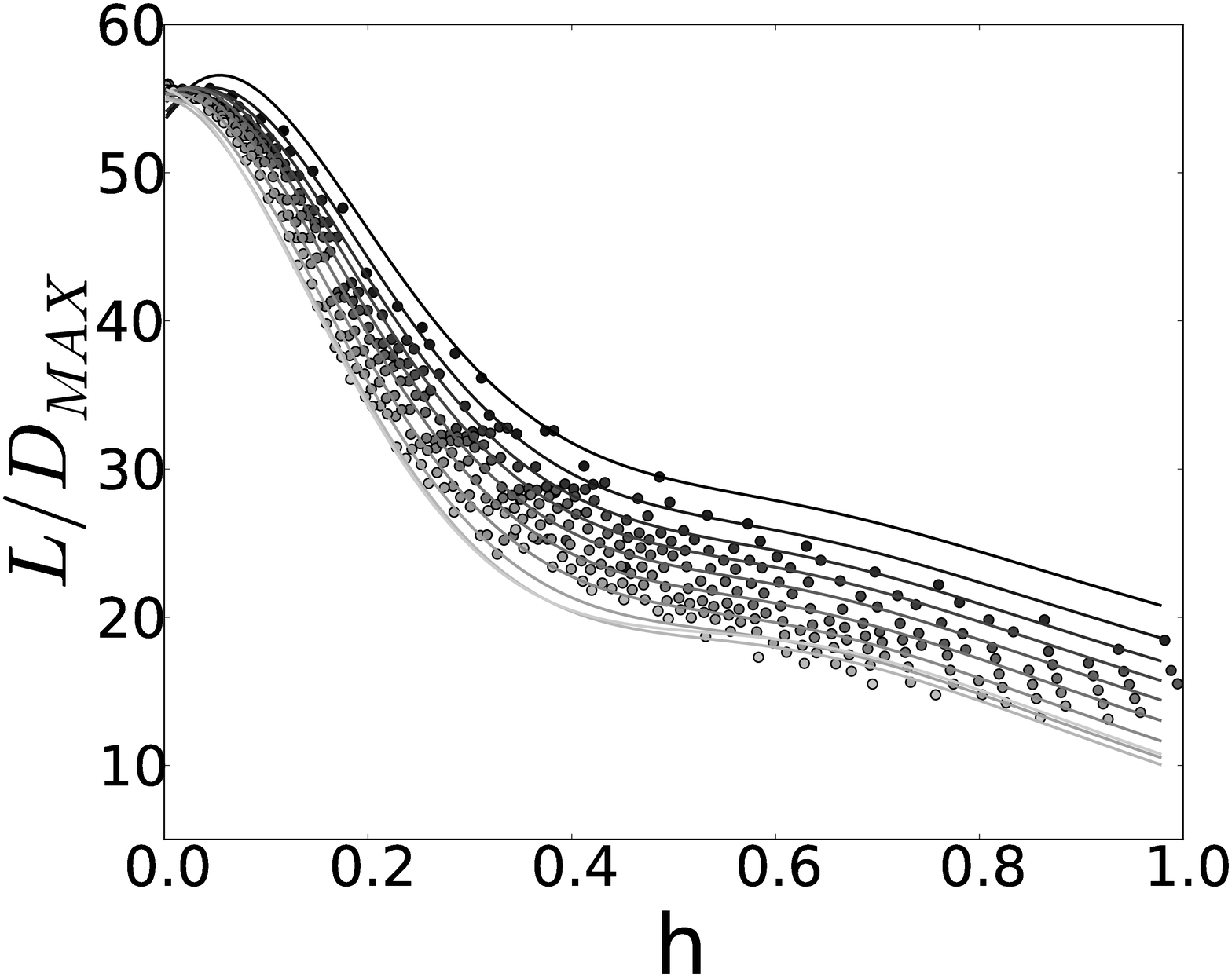}}
  \subfigure[PDF($\CLmax$)]{\includegraphics{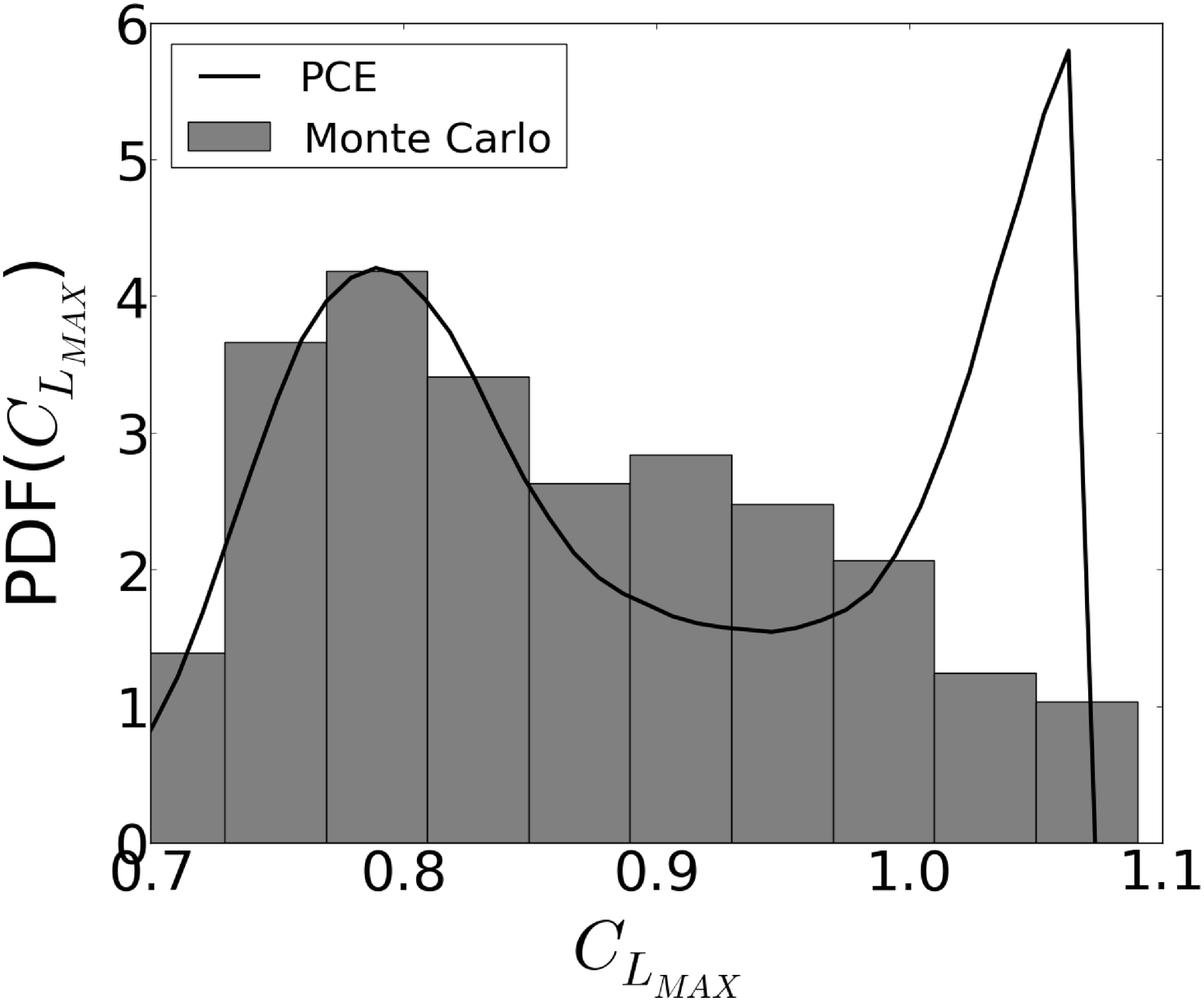}}
  \subfigure[PDF($\alphamax$)]{\includegraphics{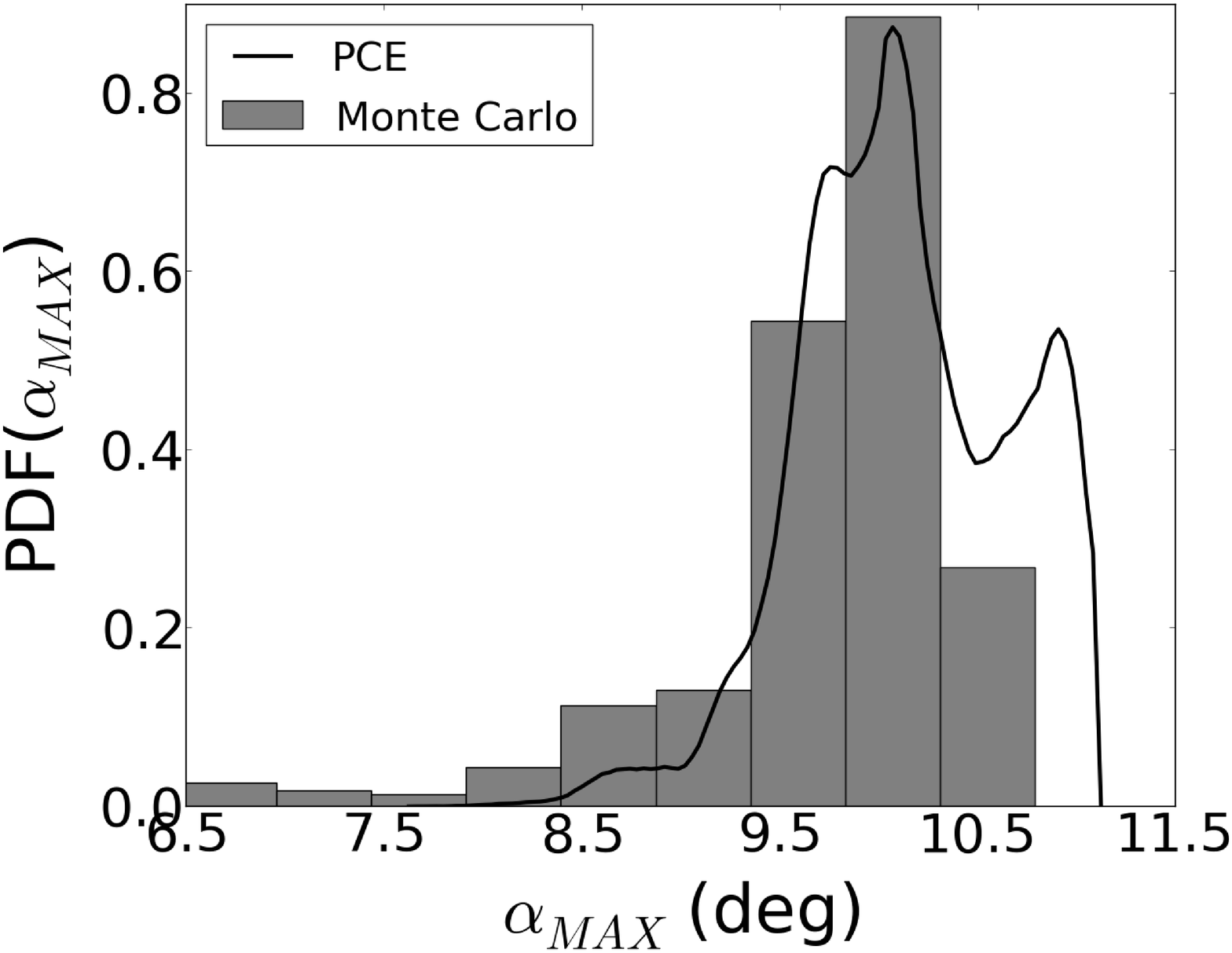}}
  \subfigure[PDF($\LDmax$)]{\includegraphics{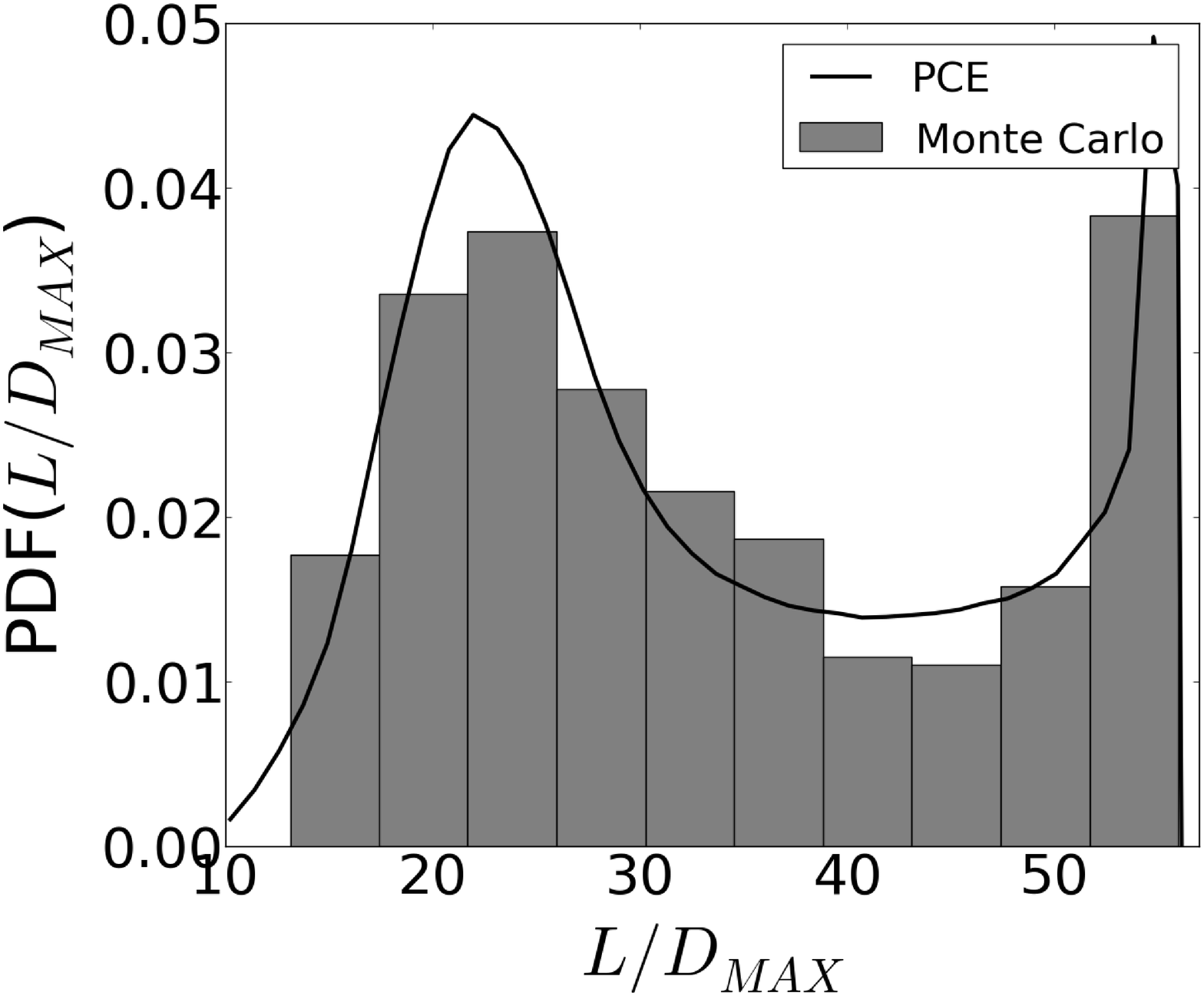}}
 \end{subfigmatrix}
 \caption{{\it TOP:} Comparisons of PCE surrogate maps to Quasi Monte Carlo
   results (the $s$ axis is going into the page). The range of
   parameter values is $h \in [0,1]$ and $s \in [0.1, 1.9]$. The grayscale is chosen to
   represent different values of $s$ (dark to light transition indicates
   increasing values of $s$). The 10 PCE curves in each plot
   represent values of $s$ equally-spaced in the interval $[0.1,1.9]$.
{\it BOTTOM:} Comparisons of the normalized PDFs
   for both the Monte Carlo and PCE cases. The input distributions used were $h$
   = $\mathcal{N}_{{1/2}^+}(0,0.5^2)$ and $s =
   \mathcal{N}(1,0.45^2)$, with both variables truncated at $2 \sigma$.}
 \label{fig:hornresults}
\end{figure}

\begin{table*}\centering
  \caption{Comparison of Statistical Moments for Monte Carlo and PCE: $h$ = $\mathcal{N}_{{1/2}^+}$(0,0.5$^2$), $s$ = $\mathcal{N}$(1,0.45$^2$)}
\ra{1}
\begin{tabular}{@{}rrrcrrcrr@{}}\toprule \toprule
& \multicolumn{2}{c}{$\CLmax$} & \phantom{abc}& \multicolumn{2}{c}{$\alphamax$ (deg)} &
\phantom{abc} & \multicolumn{2}{c}{$\LDmax$}\\
\cmidrule{2-3} \cmidrule{5-6} \cmidrule{8-9}
& MC & PCE && MC & PCE && MC & PCE \\ \midrule
Mean     & 0.86 & 0.92       && 9.7 & 10.2      && 34 & 33 \\
Variance & 0.0097 & 0.014   && 0.84 & 0.29      && 170 & 170 \\
Skewness & 0.35 & $-0.56$      && $-2.3$ & $-0.43$    && 0.38 & 0.38 \\
Kurtosis & 2.1 & 1.6         && 8.8 & 3.1       && 1.8 & 1.8 \\
\bottomrule
\end{tabular}
\end{table*}

As can be clearly seen in Fig.~\ref{fig:hornresults}, the dominant parameter is
the horn height scale ($h$); variations with horn separation distance ($s$) are
smaller by comparison. Unlike the ridge ice cases, there are discontinuities in
the maps for both $\CLmax(h,s)$ and $\alphamax(h,s)$. The PCE
expansions, which are linear combinations of smooth polynomials, obviously cannot resolve
such discontinuities. This results in errors in both the PCE surrogate maps and
in the output PDFs for $\CLmax$ and $\alphamax$. To rectify this issue,
we adopt a multi-element approach as suggested by Wan\cite{wan}.

\begin{figure}[H]
\centering
  \subfigure[$\CLmax (h,s)$]{\includegraphics[width=0.35\textwidth]{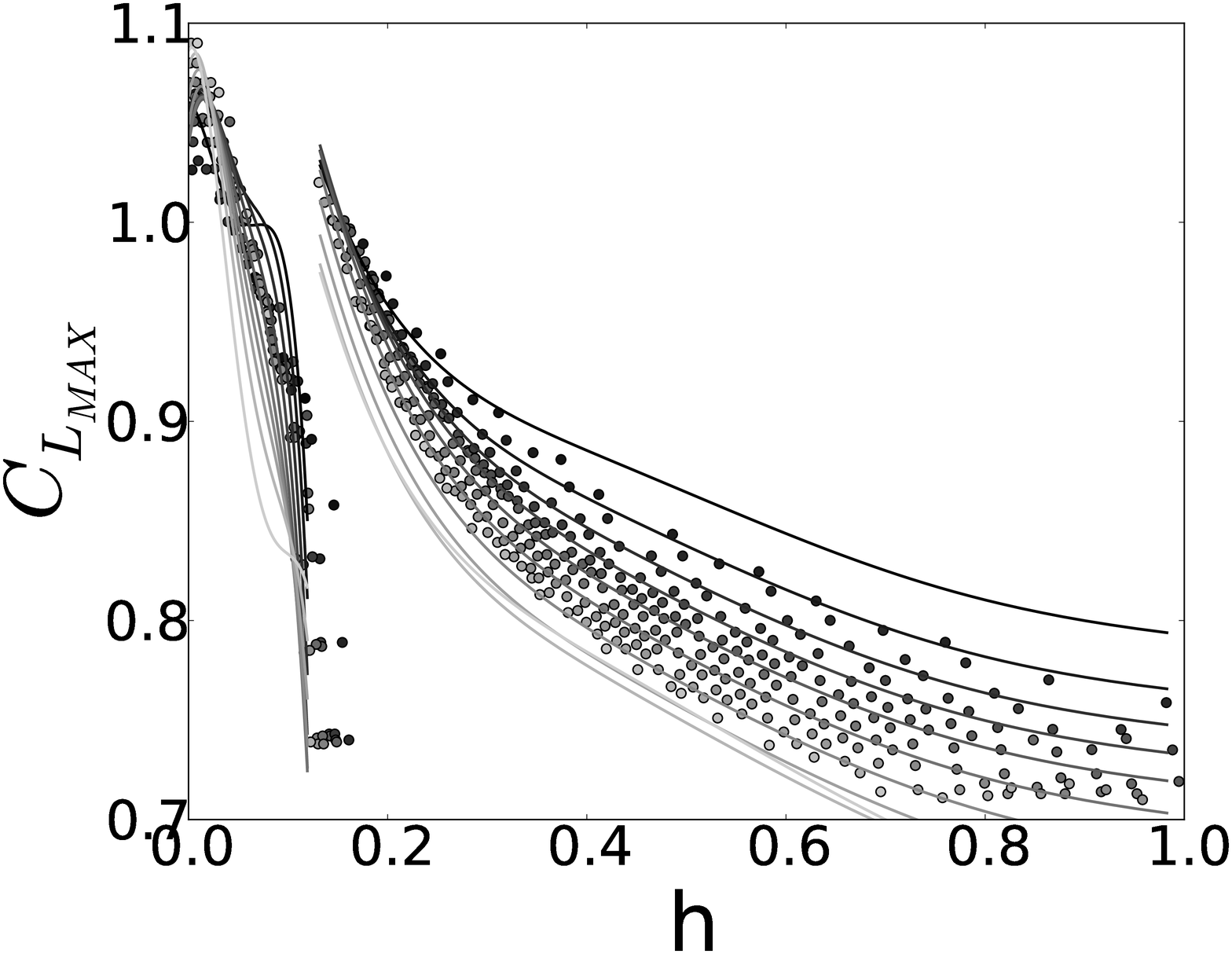}} \hspace{1 em}
  \subfigure[$\alphamax (h,s)$]{\includegraphics[width=0.35\textwidth]{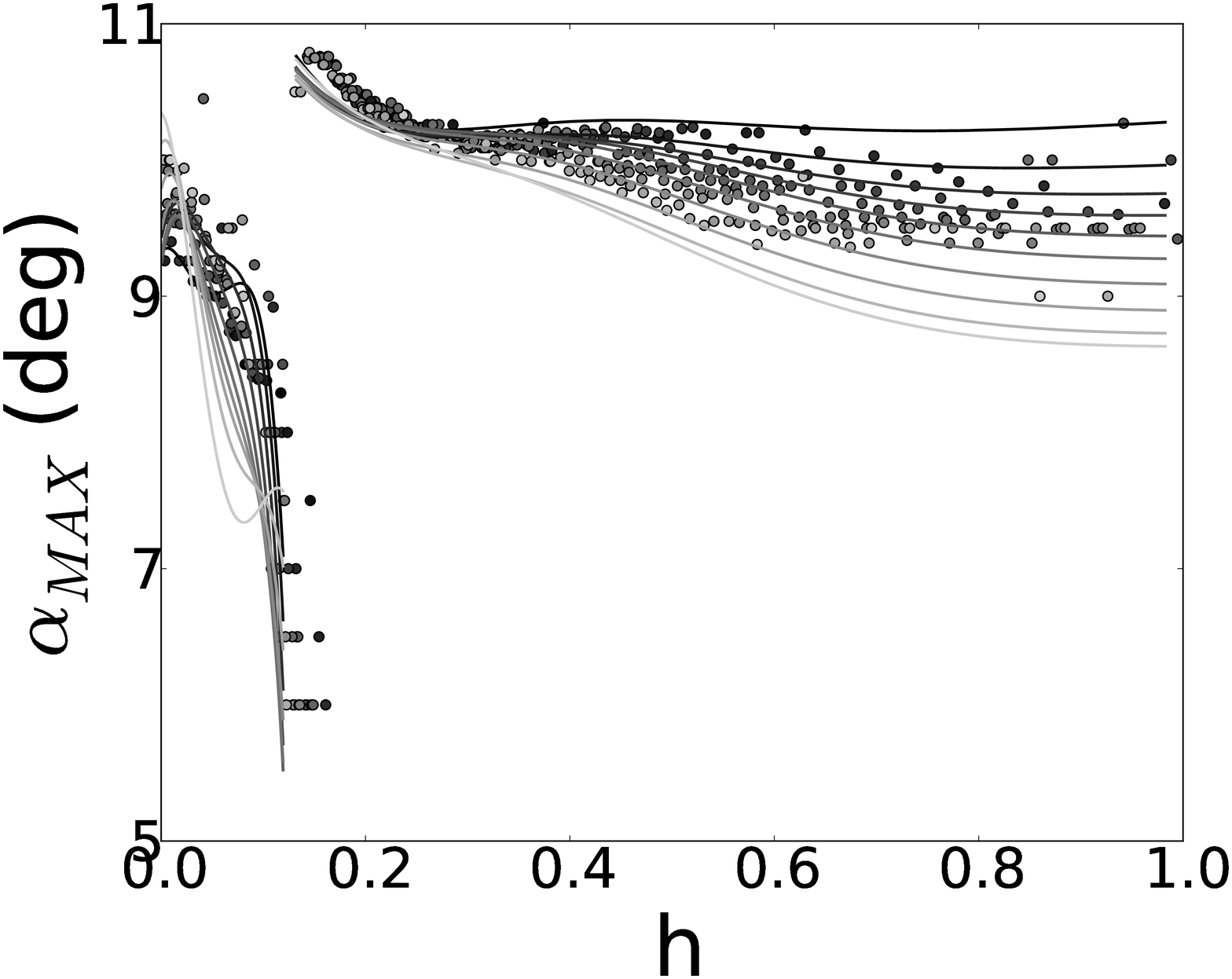}}
  \bigskip
  \subfigure[PDF($\CLmax$)]{\includegraphics[width=0.35\textwidth]{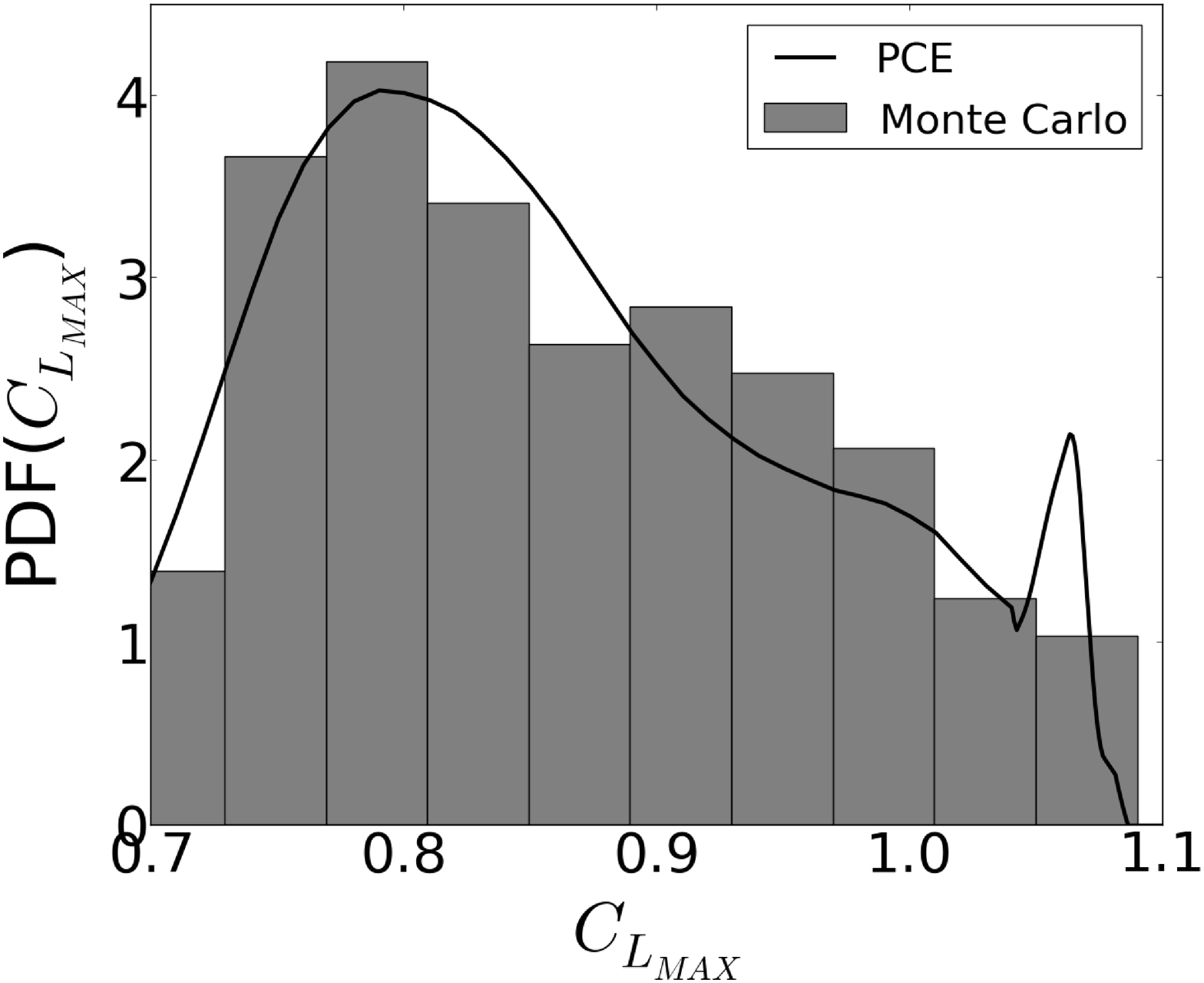}}
  \subfigure[PDF($\alphamax$)]{\includegraphics[width=0.35\textwidth]{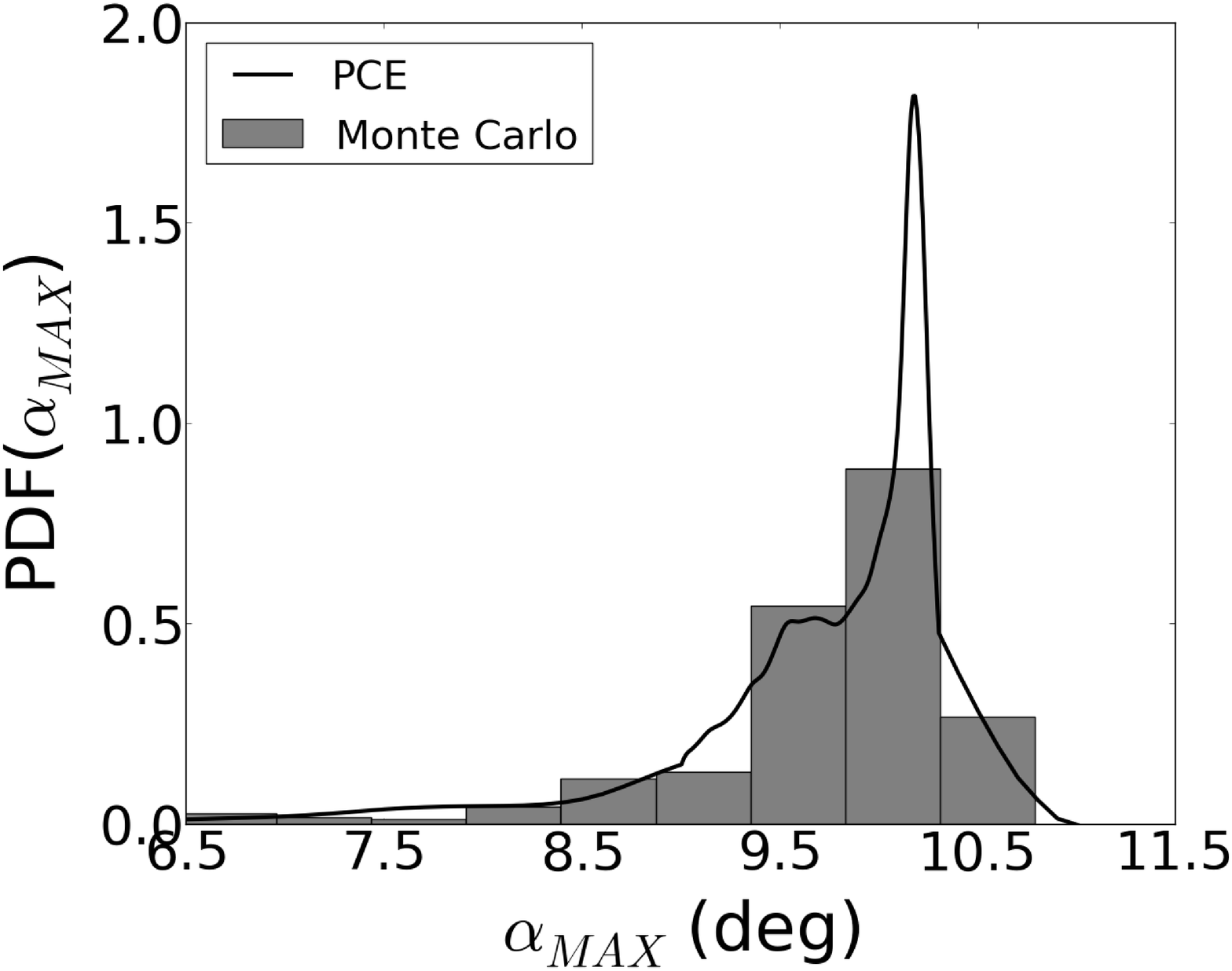}}
  \caption{{\it TOP:} Comparisons of multi-element PCE surrogate maps to Quasi
    Monte Carlo results. The range of parameter values is $h \in
    [0,1]$ and $s \in [0.1, 1.9]$. The grayscale is
    chosen to represent different values of $s$ (dark to light transition
    indicates increasing values of $s$). The 10 PCE curves in each plot
   represent values of $s$ equally-spaced in the interval $[0.1,1.9]$.
{\it BOTTOM:} Comparisons of the
    normalized PDFs for both the Monte Carlo and multi-element PCE
    cases. The input distributions used were $h$
   = $\mathcal{N}_{{1/2}^+}(0,0.5^2)$ and $s =
   \mathcal{N}(1,0.45^2)$, with both variables truncated at $2 \sigma$.}
 \label{fig:MEGPCresults}
\end{figure}

In this approach, we divide the stochastic space into two separate elements,
with the division occuring at the discontinuity ($h$-refinement). We then
implement PCE stochastic collocation separately on each of these two
elements. This results in piecewise-smooth expansions for $\CLmax(h,s)$ and
$\alphamax(h,s)$. This results in significantly better agreement between the
PCE and Monte Carlo results, as shown in Fig.~\ref{fig:MEGPCresults}.

\begin{table*}
\centering
\caption{Comparison of Statistical Moments for Monte Carlo and Multi-Element PCE: $h = \mathcal{N}_{{1/2}^+}(0,0.5^2)$, $s = \mathcal{N}(1,0.45^2)$}
\ra{1}
\begin{tabular}{@{}rrrcrr@{}}\toprule \toprule
& \multicolumn{2}{c}{$\CLmax$} & \phantom{abc}& \multicolumn{2}{c}{$\alphamax$ (deg)} \\
\cmidrule{2-3} \cmidrule{5-6}
& MC & MEPCE && MC & MEPCE \\ \midrule
Mean     & 0.86 & 0.85       && 9.7 & 9.9   \\
Variance & 0.0097 & 0.010    && 0.84 & 0.71 \\
Skewness & 0.35 & 0.50       && $-2.3$ & $-2.5$ \\
Kurtosis & 2.1 & 2.3         && 8.8 & 9.0   \\
\bottomrule
\end{tabular}
\end{table*}

\newpage

\section{Conclusions}

Wing icing is not only dangerous to pilots, it is a complex physics problem
which is subject to a large amount of uncertainty. Quantifying the exact effects
of this uncertainty on airplane performance is hence of great importance to
airplane safety.

In this work, we have demonstrated the utility of PCE methods as a fast and
accurate method for quantifying the effects on airfoil performance of ice shape
uncertainty. The main advantage of our approach is speed and
efficiency---each
of our PCE results in this paper required only 25 total simulations, compared to
the 500 simulations used in the Monte Carlo based schemes (because of the
efficiency of Gaussian quadrature). It is our hope that
improvements in icing UQ can contribute towards improved safety regulations and
protocols for pilots and a mitigation of icing-related accidents.

\section*{Acknowledgments}
These authors would like to acknowledge the FAA Joint University Program for Air
Transportation (JUP) for their support of this research.

\bibliographystyle{aiaa}	
\bibliography{GPCUQREF}		

\end{document}